# Energy-based time derivative damage accumulation model under uniaxial and multiaxial random loadings

Shih-Chuan Tien, Haoyang Wei, Jie Chen, Yongming Liu*

*School for Engineering of Matter, Transport, and Energy, Arizona State University, Tempe, AZ 85287, USA*

## ABSTRACT:

A new fatigue life prediction method using the energy-based approach under uniaxial and multiaxial random loadings is proposed in this paper. One unique characteristic of the proposed method is that it uses time-derivative damage accumulation model compared to the classical cycle-based damage accumulation model. Thus, damage under arbitrary random loading can be directly obtained using time-domain integration without cycle counting (e.g., rain-flow cycle counting method in classical fatigue analysis). First, a brief review of existing models is given focusing on their applicability to uniaxial/multiaxial, constant/random, and high cycle fatigue/low cycle fatigue loading regimes. It is observed that most existing models are only applicable to certain loading conditions and many of them are not applicable/validated under random loadings. Next, a time-derivative damage accumulation model under uniaxial random loading is proposed. The proposed damage function is inspired by a time-domain fatigue crack growth model. The fatigue life is obtained by integrating the damage function following random energy loading histories. Following this, an equivalent energy concept for general multiaxial loading conditions is used to convert the random multiaxial loading to an equivalent random uniaxial loading, where the time-derivative damage model can be used. Finally, the proposed model is validated with extensive experimental data from open literature and in-house testing data under various constant and random spectrum loadings. Conclusions and future work are suggested based on the findings from this study.

* Corresponding author
Email: yongming.liu@asu.edu
Phone: 480-965-6883

## 1 Introduction

Fatigue is a critical issue that has been researched for many years. An accurate prediction of fatigue life is very important to ensure the integrity of structural component used in the real world. Theoretically speaking, all structural components are under general multiaxial random loadings, although simplifications have been made to approximate the problem as uniaxial and/or constant amplitude loadings. Existing fatigue models for general multiaxial fatigue loadings can be classified as stress-based [1–5], strain-based [6–8], energy-based [9–13], fracture mechanics-based [14,15], and other methods [16,17]. A brief review is given below, and the focus is on the applicability of existing models for different loading conditions, such as uniaxial vs. multiaxial, constant vs. random, and high cycle fatigue vs. low cycle fatigue.

The stress-based approach is a widely used method used to predict the fatigue life, especially in high cycle fatigue (HCF) regime. For example, the stress-based model, proposed by Liu and Mahadevan [5], modified the critical plane approach to develop a new criterion to predict fatigue life in multiaxial constant amplitude loading. Wei and Liu [2] extended the modified critical plane approach [5] and combined the rainflow-counting algorithm and Miner's rule to develop a new model that is suitable for predicting fatigue lives under multiaxial random loadings. Other stress-based approaches have been proposed to use different high-cycle fatigue criteria to predict in-phase and out-of-phase multiaxial fatigue [18–20] and other aspects such as predicting the fatigue life by using probability method [21–23]. However, the stress-based model has a drawback such that the model is not suitable to deal with the low cycle fatigue (LCF) regime. When the material deformation approaches the plastic region, the strain-based approach is more appropriate.

The strain-based approach was proposed by several authors based on different concepts. For example, a strain-based model [6] is based on the characteristic plane approach and includes an empirical non-proportional hardening factor. Another strain-based model proposed by Remes et. al. [7] is developed to predict the fatigue life of a welded steel joint considering notch effect. Since the strain-based method is only related to strain, the out-of-phase hardening phenomenon is difficult to be handled. Thus, energy-based model, where both stress and strain are used, can intrinsically consider the out-of-phase hardening phenomenon.

The energy-based approach has been proposed [10–12,24,25] to predict both HCF and LCF fatigue lives. The energy-based approach directly applies the stress and strain history spectrum to acquire energy spectrum to calculate the fatigue life. Energy-based approach takes material constitutive response into account and can directly correlate energy with fatigue damage. Therefore, the energy-based model is chosen in this study.

Beyond the three classifications above, there are a few different approaches to predict the fatigue life. The fracture mechanics-based approach based on the fatigue crack growth analysis and the equivalent initial flaw size (EIFS) concept has been proposed for fatigue life prediction [26]. The fatigue model with the EIFS concept is suitable for both uniaxial and multiaxial constant loading cases [26,27]. No demonstration and validation for random loading cases were provided. Multiaxial constant loading cases have been predicted very well by several other models [1,5,6,11]. Again, the demonstration and validation of these models for random loadings is lacking.

The random fatigue loading spectrum includes various peaks/valleys and sometimes are not even cyclic (e.g., Christmas tree-type spectrums). In practice, cycle counting algorithms (e.g., rainflow-counting algorithm) and a damage accumulation rule (e.g., Miner's rule) are combined with the multiaxial fatigue model for fatigue life prediction under random loading spectrums. A review that summarizes several criterions for solving random loading cases can be found in [28]. A detailed comparison for different models reviewed here is shown in Table 1. Each different model solves different loading cases and conditions. Detailed model formulation can be found in the corresponding reference. It should be noted that this is not a complete list of all available models in the open literature and only shows some representative categories.

Table 1 Comparison of different fatigue life predicted methods

| Methods | Loading spectrum | | Loading path | | Loading regime | | Ref. |
|---|---|---|---|---|---|---|---|
| | Constant | Random | Uniaxial | Multiaxial | HCF | LCF | |
| Stress-based model | √ | | √ | √ | √ | | [1] |
| Stress-based model | √ | √ | √ | √ | √ | | [2] |
| Stress-based model | √ | √ | √ | √ | √ | | [3] |
| Stress-based model | √ | | √ | √ | √ | | [5] |
| Strain-based model | √ | | √ | √ | √ | √ | [6] |
| Strain-based model | √ | | √ | √ | √ | √ | [29] |
| Strain-based model | √ | | √ | √ | √ | √ | [30] |
| Energy-based model | √ | | √ | | √ | √ | [10] |
| Energy-based model | √ | | √ | √ | √ | √ | [11] |
| Energy-based model | √ | | √ | √ | √ | √ | [12] |
| Energy-based model | √ | | √ | √ | √ | √ | [13] |
| Energy-based model | √ | | √ | √ | √ | √ | [25] |
| Crack growth-based model (EIFS) | √ | | √ | √ | √ | √ | [26,27] |
| Energy-based model | √ | | √ | √ | √ | √ | [31] |
| Energy-based model | √ | | √ | √ | √ | √ | [32] |
| Continuum damage mechanics method | √ | | √ | √ | √ | √ | [33] |
| Time-derivative model | √ | √ | √ | √ | √ | √ | Proposed |

In Table 1, most of methods do not cover all loading conditions. This is especially true for random loading spectrums. In view of this, the motivation of this study is to develop a model which can handle all these different loading conditions, especially for random loadings. It is also noticed that all these reviewed methods are cycle-based. The cycle-based method relies on the cycle-counting algorithm and cycle-based damage accumulation rule, which is very difficult for general multiaxial random loadings. For example, critical plane-type approach needs to know the critical plane orientation, which is not known as a priori for random loadings. Some averaged calculation is needed. In addition, cycle counting results based on different load channels (e.g., tensile or tortional loading) might be totally different if the loading frequency differs. Finally, cycle-counting method ignores the sequence effect in the true time-history. Thus, it is intrinsically difficult if nonlinear damage accumulation occurs. In view of this, the proposed study proposes to use an alternative approach for fatigue analysis which is based on the time-derivative definition of damage. Time-based damage concept can track every damage increment under every time step. The damage accumulation can be in analogy to a time-domain integration and avoids the ambiguity of cycle-counting and cycle definition.

The remaining of this paper is organized as follows. The time-derivative model under uniaxial loading will be introduced first. A new damage accumulation function in analogy to fatigue crack propagation will be discussed in detail. Next, the time-based concept is integrated with an equivalent energy concept to extend the model to general multiaxial random loadings. Following this, the validation of the model using several databases from open literature and in-house testing will be presented. Finally, a conclusion and discussion of the results will be presented. The future work is also suggested in the end.

## 2 Proposed Methodology

### 2.1 Time-derivative damage accumulation model under uniaxial loading

The time-derivative concept is illustrated in Fig. 1. Fig. 1 shows the relationship between the strain energy and time for an arbitrary uniaxial random loading spectrum. The shaded area corresponds to the damage accumulation happening in the loading phase. The damage value in every time point can be calculated directly by integrating with respect to time. Due to the energy requirements, no damage accumulation will happen in the unloading phase. Classical damage accumulation in fatigue analysis is cycle-based and is usually associated with cycle-counting algorithms (e.g., rain-flow cycle counting algorithms) for random loadings. One unique difference of the proposed time-derivative damage accumulation model is that it automatically includes the sequence effect while the flow-accounting algorithms will lose the sequence information. Another major difference is that the cycle-based damage accumulation is for the average behavior per cycle and did not specify the damage accumulation rate at the initial or at the end of the loading cycle. The time-derivative damage model will specify the instantaneous damage accumulation rate at every time point (i.e., load point). Therefore, the fatigue life can be directly predicted by integrating the damage with respect to time (alternatively with respect to load history).

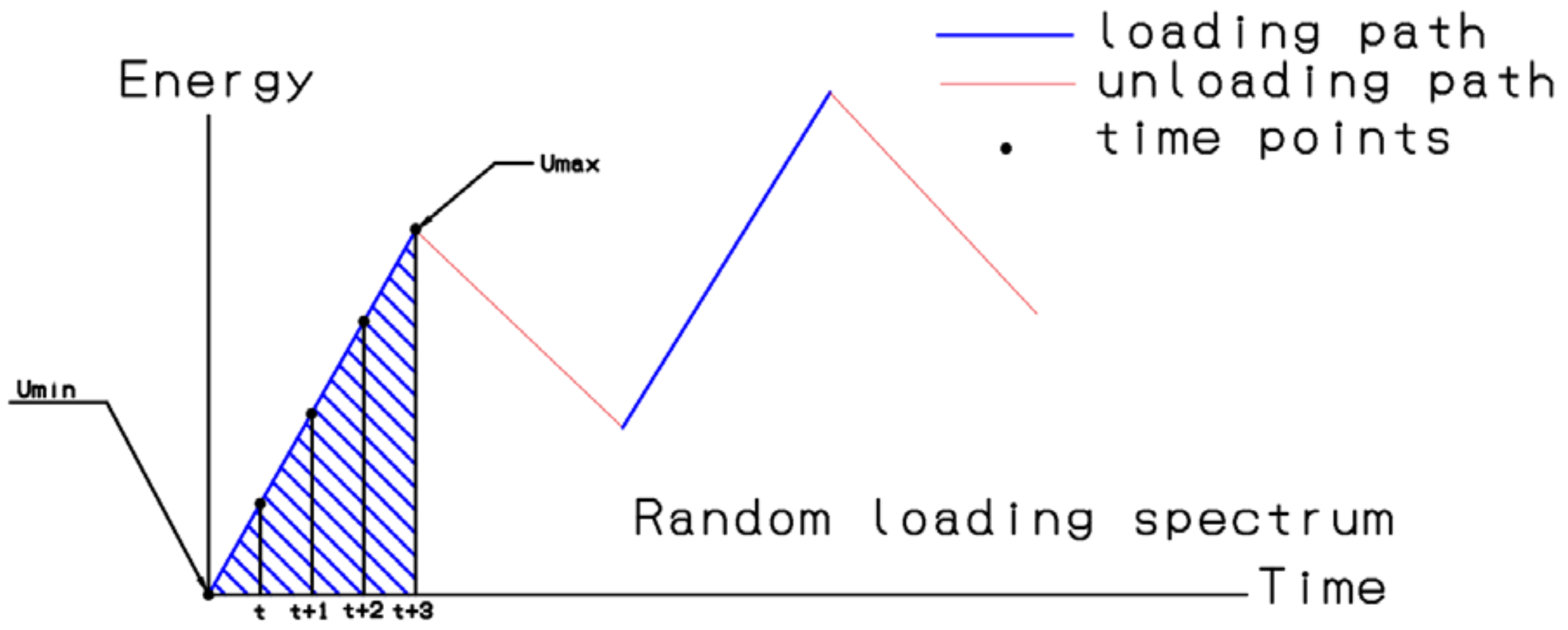

Fig. 1 Schematic illustration of the time-derivative damage model

The remaining question is to define the damage and damage rate for this time-derivative model. Classical cycle-based damage accumulation model defines the damage from the "cycle fraction" concept, which is not applicable for time-derivative model. The proposed time-derivative concept uses the analogy approach with respect to a time-derivative fatigue crack growth (FCG) function, proposed in [34]. The key concept is to treat the damage like a crack in the FCG analysis and assumes the same function type for the rate function and integral. In the time-derivative FCG analysis, the crack length can be calculated by integrating the instantaneous crack growth function with respect to time/loading spectrum. The classical crack growth rate per cycle can be predicted as

$$\frac{da}{dN} = \int_{\sigma_{min}}^{\sigma_{max}} \frac{da}{d\sigma}\left(\sigma, \dot{\sigma}, a, E, \sigma_y \ldots\right) d\sigma \quad (1)$$

where $\frac{da}{dN}$ is crack growth per cycle in classical FCG anlysis. $\sigma_{max}$ and $\sigma_{min}$ are maximum and minimum stress, respectively. $\frac{da}{d\sigma}\left(\sigma, \dot{\sigma}, a, E, \sigma_y \ldots\right)$ is a time-based crack growth function that depends on stress $\sigma$, loading rate $\dot{\sigma}$, crack length $a$, Young's modulus of material $E$, yield strength $\sigma_y$, and other parameters. Eq. (1) represents the generic expression for the time-derivative FCG and many different function form of $\frac{da}{d\sigma}\left(\sigma, \dot{\sigma}, a, E, \sigma_y \ldots\right)$ can be used.

The proposed study follows the definition based on the geometrical relationship between incremental crack growth $da$ and the change in the crack tip opening displacement (CTOD) [35]. The details are not discussed here, and readers can refer to the cited articles using in situ experimental imaging analysis [36,37]. The relationship between crack length increment and CTOD is expressed as

$$\Delta a = A K_{max}^{B} \delta^{d} \quad (2)$$

where $\Delta a$ is the incremental crack length in the current cycle. $A$ , $B$ and $d$ are fitting parameters. $K_{max}$ is the maximum stress intensity factor from the previous loading cycle. $\delta$ is the CTOD. A schematic plot of this function is shown in Fig. 2. In the experiment, it is observed that the higher $K_{max}$ usually leads to higher crack growth rate as more damage is introduced by the high maximum loading (e.g., more slip plans and more micro cracks ahead of the crack tip). Thus, the curve with a higher $K_{max}$ tends to have a faster crack growth rate. Mathematically, $B$ is a positive value to reflect this trend. In addition, it is observed that the crack growth rate is higher for very small CTOD values and is slowing down for higher CTOD values. The reason is that the crack is very sharp when CTOD is small and tends to become a blunted notch when the loading increases.

This is due to the large plastic deformation near the crack tip and reduces the stress intensity. Thus, crack growth rates slow down. Mathematically, $d$ should be a value less than 1.

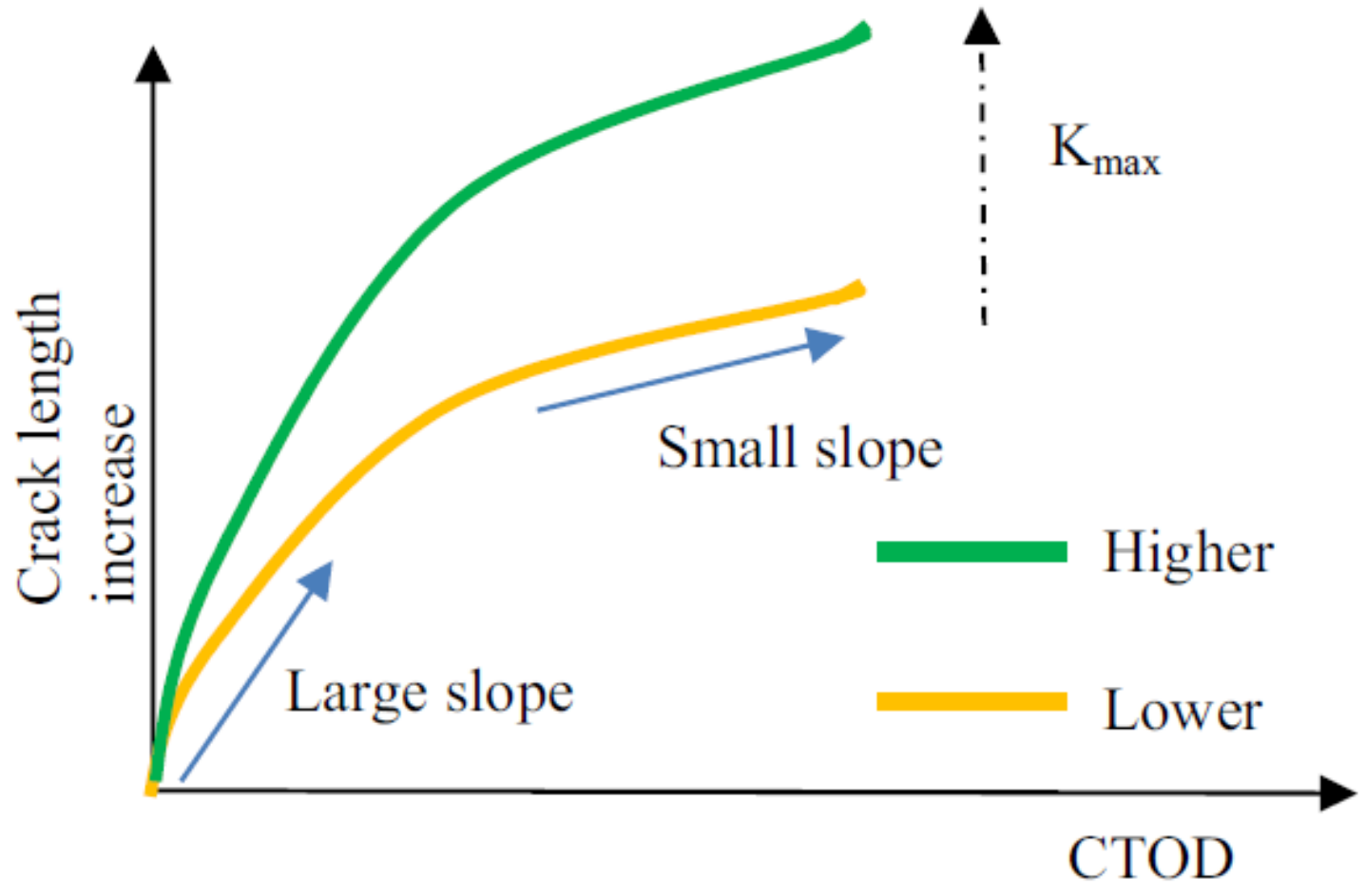

Fig. 2 The Relationship of Crack Length Increase Rate and CTOD at Sub-Cycle Scale [34]

The proposed study assumes that the damage accumulation is analogy to the crack growth, in terms of their function type and integration strategy. The proposed time-derivative damage growth function is similar to the crack growth function, but replaces crack length with damage and replaces the stress with energy. Thus, the time-derivative damage model can be expressed as

$$\frac{dD}{dN} = \int_{U_{min}}^{U_{max}} \frac{dD}{dU}\left(U, \dot{U}, D, E, \sigma_y \dots\right) dU \tag{3}$$

where $\frac{dD}{dN}$ is damage growth per cycle. $U_{max}$ and $U_{min}$ is the maximum energy and minimum energy, respectively. The $\frac{dD}{dU}\left(U, \dot{U}, D, E, \sigma_y \dots\right)$ is a time-based damage growth function that depends on the energy $U$, energy rate $\dot{U}$, damage $D$, Young's modulus of material $E$, yield strength $\sigma_y$, and other parameters. Next, the relationship between damage and CTOD can be shown as

$$\Delta D = AK'^{B}_{max}\delta'^{d}, \tag{4}$$

where $\Delta D$ is incremental damage in the current cycle, and the other parameters are the same as those in Eq. (2). It should be noted that we use $K'$ and $\delta'$ to replace $K$ and $\delta$ because the stress intensity factor $K$ and CTOD $\delta$ are no longer valid as there is no crack. Instead, all terms of crack length $a$ is replaced with the damage variable $D$. For example,

$$K_{max} = \sigma(\pi a)^{\frac{1}{2}} \quad \rightarrow \; K'_{max} = \sigma(\pi D)^{\frac{1}{2}} \tag{5}$$

The proposed model is energy-based and the stress is replaced by energy. For ideal elastic response, we have

$$\sigma = \sqrt{2EU} \tag{6}$$

where $E$ is the Young's modulus, $U$ is the strain energy, and $\sigma$ is the stress. Substituting Eq. (6) to Eq. (5), one can obtain

$$K'_{max} = (2\pi DEU_{max})^{\frac{1}{2}} \tag{7}$$

The same modification can be applied to CTOD $\delta$ as well. Taking the differential form for both sides of Eq. (4), the equation can be expressed as

$$\frac{dD}{dt} = AK'^{B}_{max}\delta'^{d-1}\frac{d\delta\prime}{dt} \tag{8}$$

where $\frac{dD}{dt}$ is the time-derivative damage growth rate. It should be noted that $K'_{max}$ refers to the largest value from the previous loading history and is considered to be a constant in the current loading cycle. Thus, only $\delta'$ is considered for the derivative operation. Eq. (8) is the proposed time-derivative damage model. If material response is not rate-dependent (e.g., no creep, corrosion, oxidation damage), fatigue damage will only depend on the load variation. Thus, $dt$ terms in Eq. (8) can be cancelled and the time-deravative function can be expressed as energy-deravative function. Details are shown below.

Eq. (4) can be rewritten by expanding all terms using the damage variable $D$ and the energy variable $U$ as

$$\Delta D = A(2\pi DEU_{max})^{\frac{B}{2}}(\frac{2\pi DEU}{2E\sigma_{y}})^{d} \tag{9}$$

where $\Delta D$ is the incremental damage. Energy-derivative formulation can be rewritten as

$$\frac{dD}{dU} = Ad(2\pi E U_{max})^{\frac{B}{2}} (\frac{\pi}{\sigma_y})^d D^{\frac{B}{2}+d} U^{d-1} \tag{10}$$

Substituting Eq. (10) to Eq. (3), one can obtain

$$\frac{dD}{dN} = 2Ad(2\pi E U_{max}^{his})^{\frac{B}{2}} (\frac{\pi}{\sigma_y})^d D^{\frac{B}{2}+d} \int_{U_{min}}^{U_{max}} U^{d-1}\, dU \tag{11}$$

where $\frac{dD}{dN}$ is damage increment per cycle. $\int_{U_{min}}^{U_{max}} U^{d-1}\, dU$ refers to the accumulated energy in the current completed cycle. $U_{max}^{his}$ is a memory energy value and depends on the loading history. For constant loading, the value is the same as the maximum energy $U_{max}$ in the equation. For random loading, $U_{max}^{his}$ may come from an earlier cycle due to the memory effect. In the current study, a memory window is selected and the maximum value from this memory window is used for damage calculation. Various memory window size can be selected, and the current study uses the value of one hundred complete load reversals before the current calculation. The experimental validation later in this paper shows great accuracy for this memory window size. As can be seen in Eq. (11), the damage can be continuously accumulated by direct numerical integration with respect to the energy loading history. The proposed method does not need cycle counting by definition and Eq. (11) is application for both constant and random loadings. Thus, we do not need separate models for constant amplitude loading and variable amplitude loading, as typically seen in classical cycle-based fatigue models.

Next question is the fatigue failure criteria and we followed the same concept as the fatigue damage accumulation: the material will fail when the value of accumulated damage exceeds 1. Eq. (11) is simplified for easy presentation by introducing an α to make the equation more concise. Eq. (11) is rewritten as

$$\frac{dD}{dN} = 2d\alpha D^{\frac{B}{2}+d} \int_{U_{min}}^{U_{max}} U^{d-1}\, dU, \tag{12}$$

where $\alpha$ is expressed as

$$\alpha = A(2\pi E U_{max}^{his})^{\frac{B}{2}} (\frac{\pi}{\sigma_y})^d. \tag{13}$$

Eq. (12) is further simplified by integrating with respect to the energy and rearrange it, the equation can be shown as

$$\frac{dD}{dN} = \alpha D^{-(\frac{B}{2}+d)}(U_{max}^{d} - U_{min}^{d}). \tag{14}$$

Now, the equation is related to maximum energy and minimum energy if the loading is a complete cycle. Eq. (12) can also be expressed if the loading is for two continuous time point as

$$dD = \alpha D^{-(\frac{B}{2}+d)}(U_{t+1}^{d} - U_{t}^{d}), \tag{15}$$

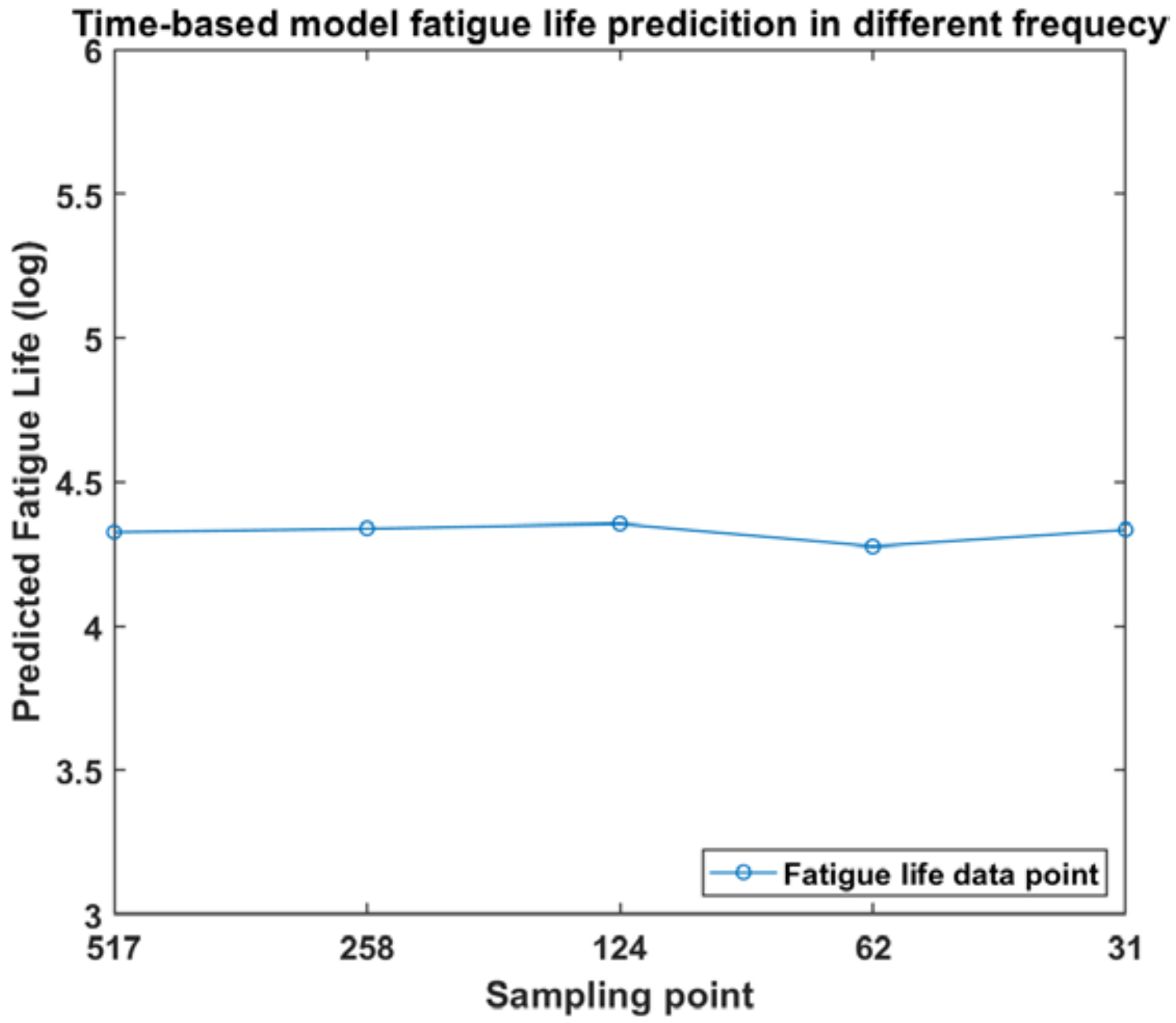

Fig. 3 The Predicted Fatigue Life in Different Sampling Rate

where $U_{t+1}$ is the energy value at the next point and $U_t$ is the energy value at the current point. It should be noted that Eq. (14) and Eq. (15) are identical if the minimum load happens at the time $t$ and the maximum loading happens at the time $t+1$. In practice, either one can be used and Eq. (15) offers slight benefits as it does not require the identification of peak and valley. The final life prediction is independent of the sampling frequency. A numerical verification example is shown in Fig. 3. The sampling point per cycle is varied from 31 to 517 and the life prediction results are plotted. It is observed that almost same life prediction results are obtained. The results converged when the number of sampling point (e.g., sampling frequency) increases. A slight variation is observed for lower sampling frequency. The reason is that a lower sampling frequency may miss the actual peak and valley values and causes numerical errors.

Finally, the fatigue life can be calculated directly by checking to see if the accumulated damage is larger than 1, i.e.,

$$D_f = D_{t-1} + dD \geq 1, \tag{16}$$

where $D_f$ is the total damage at the failure, $D_{t-1}$ is the cumulative damage in the previous time, and $dD$ is the incremental damage in the current time point. Eqs. (15-16) are the proposed time-derivative damage accumulation model and failure criteria under uniaxial loadings. Similar with many other fatigue models, there are some material parameters in the proposed study that need to be determined. As it has been shown that the proposed methodology works for both constant and random loadings, experimental data under arbitrary loading can be used for calibration. For convenience, we used constant amplitude loading S-N curve data for calibration, which has extensive database from open literatures. This also eases the future application of the proposed methodology as the time-derivative formulation is relatively new in the fatigue community. Detailed calibration procedure is shown below.

## 2.2 Model Parameter Calibration Using Constant Amplitude S-N Curve Data

This section tries to calibrate the proposed model parameters using widely available constant amplitude S-N curve data. Analytical solution is obtained for easy implementation in the future. The proposed study used the energy instead of stress/strain in the classical S-N curves. The energy-curve under uniaxial constant amplitude loading is assumed to follow the classical Basquin equation (i.e., power function) as

$$U = p \times {N_f}^q, \tag{17}$$

where $U$ is the energy range. $N_f$ is the fatigue life from the experimental data. $p$, $q$ are fitting coefficients. It has been shown in the last section that the fatigue life can be predicted by integrating the damage growth function. If we use the similar notation in the classical Paris's law as

$$\frac{da}{dN} = C(\Delta K)^m \qquad \rightarrow \qquad \frac{dD}{dN} = C(\Delta K')^m \;= AK'^{B}_{max}\delta'^{d}. \tag{18}$$

From classical FCG analysis [38], the fitting parameters $C$ and $m$ can be uniquely solved by mapping to the $p$ and $q$ in fatigue S-N curve as

$$\begin{cases} C = \frac{2}{(2-m)}(1 - D_0^{\frac{2-m}{2}})(2\pi Ep)^{-\frac{m}{2}} \\ m = -\frac{1}{q} \end{cases}, \tag{19}$$

where $D_0$ is the initial damage in the material. In the current study, initial damage is assumed a very small value and can be approximated as zero. By comparing Eq. (18) and Eq. (19), one can have

$$\begin{cases} A = C(0.36)^{-d}(1-R)^B(2E\sigma_y)^d \\ B = m - 2d \end{cases} \text{for} \quad R>0 \tag{20}$$

where R is the stress ratio from the constant amplitude fatigue testing.

For R < 0, a modification factor is added as [34]

$$\begin{cases} A = C(0.36)^{-d}(1-R)^{B+2d}(1-\beta R)^{-2d}(2E\sigma_y)^d \\ B = m - 2d \end{cases} \quad \text{for R <0} \tag{21}$$

β is calibrated with a negative stress ratio fatigue crack growth testing. Details about β can be found in [34]. It varies for different metal materials. An approximation is given in

$$\beta = 30.091\sigma_y^{-0.797}. \tag{22}$$

It is seen that the proposed method has three parameters that need to be calibrated: *A*, *B*, and $d$. Fatigue S-N curve has two independent fitting coefficients: $p$ and $q$. No unique solution for *A*, *B*, and $d$ can be obtained. In the proposed study, we fixed the value of $d = 0.01$ to get feasible parameter calibration for fatigue life prediction. The authors tried different $d$ values and experimental validation shows that $d = 0.01$ works well for the investigated loading and materials.

## 2.3 Extension to Multiaxial Loading using Equivalent Energy Concept

The above discussion is for uniaxial random loadings. The section focuses on the extension to general multiaxial random loadings. An equivalent energy concept for multiaxial constant amplitude loading was proposed by Wei and Liu [11]. The key concept is to convert the multiaxial loading to an equivalent uniaxial tension-compression loading. Thus, uniaxial fatigue energy-curves can be used for life prediction. The proposed model uses the same concept and tries to convert the multiaxial random loading to an equivalent uniaxial random loading, which can employ the methods in the previous sections for life prediction. In this concept, energy is classified into two major components: dilatational and distortional energy, which is calculated under general three-dimensional conditions. The dilatational energy represents the volumetric change, and the distortional energy represents the shape change in the material. Thus, the total energy is expressed as

$$U_{total} = U_{dil} + U_{dis} \tag{23}$$

where $U_{total}$ is the total energy, $U_{dil}$ is the dilatational energy, and $U_{dis}$ is the distortional energy. The dilatational and distortional energy can be expressed using three other energy terms: tensile energy, torsional energy, and hydrostatic energy. The final equation can be expressed as

$$U_{ten}^{eq} = U_{ten} + sU_{tor} + kU_H \tag{28}$$

where $U_{ten}^{eq}$ is the equivalent energy. $U_{ten}$ is the tensile energy. $U_{tor}$ is the torsional energy. $U_H$ is the hydrostatic energy. A detailed description of the method can be found in [11]. A schematic illustration of the conversion process is shown in Fig. 4.

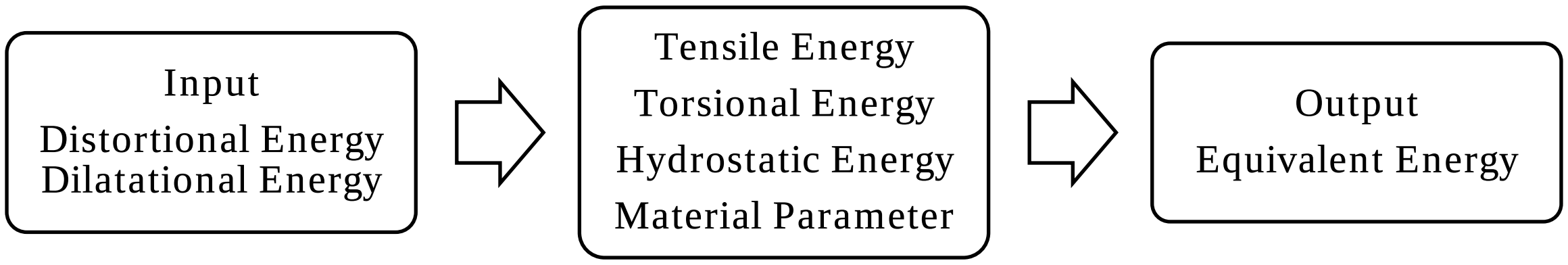

Fig. 4 Schematic Illustration of Equivalent Energy Conversion Process

## 3 Model Validation

The proposed model is validated with experimental data from several open literatures [2,29,30,33,39–43] and from in-house fatigue testing spectrums. The datasets include uniaxial and multiaxial loading. Most literature data are only for constant amplitude loading case. Thus, in-house testing is performed to collect data under general random variable spectrum loadings. Data collection and generation details are shown below.

### 3.1 Data collection from open literature and model validation

Table 2 shows the summary of the materials and several loading paths. The references are listed in the last column and the material properties can be found in the references. The collected data includes several load paths (see names in Table 2) and are plotted in Fig. 5. The results of the predicted fatigue life and experimental fatigue life under constant loading for seven materials are shown in Fig. 6. As can be seen, most of the model predictions are within a life factor 2-3, which is considered as very good accuracy in fatigue.

Table 2. Summary of Collected Experimental Data.

| Material | Loading Path | Reference |
|---|---|---|
| AISI 304 Steel | Uni, Tor, Pro, Sin90 | [29] |
| A533B | Uni, Tor, Pro, Sin90 | [39] |
| S45C Steel | Uni, Tor, Pro, Sin90, Sin45, Sin22.5 | [30] |
| SAE 1045 | Uni, Tor, Pro, Sin90, box | [40] |
| S460N | Uni, Tor, Pro, Sin90 | [41] |
| Al T6061 | Uni, Tor, Sin90 | [42] |
| Al T7075 | Uni, Tor, Pro, Sin90 | [2,43] |

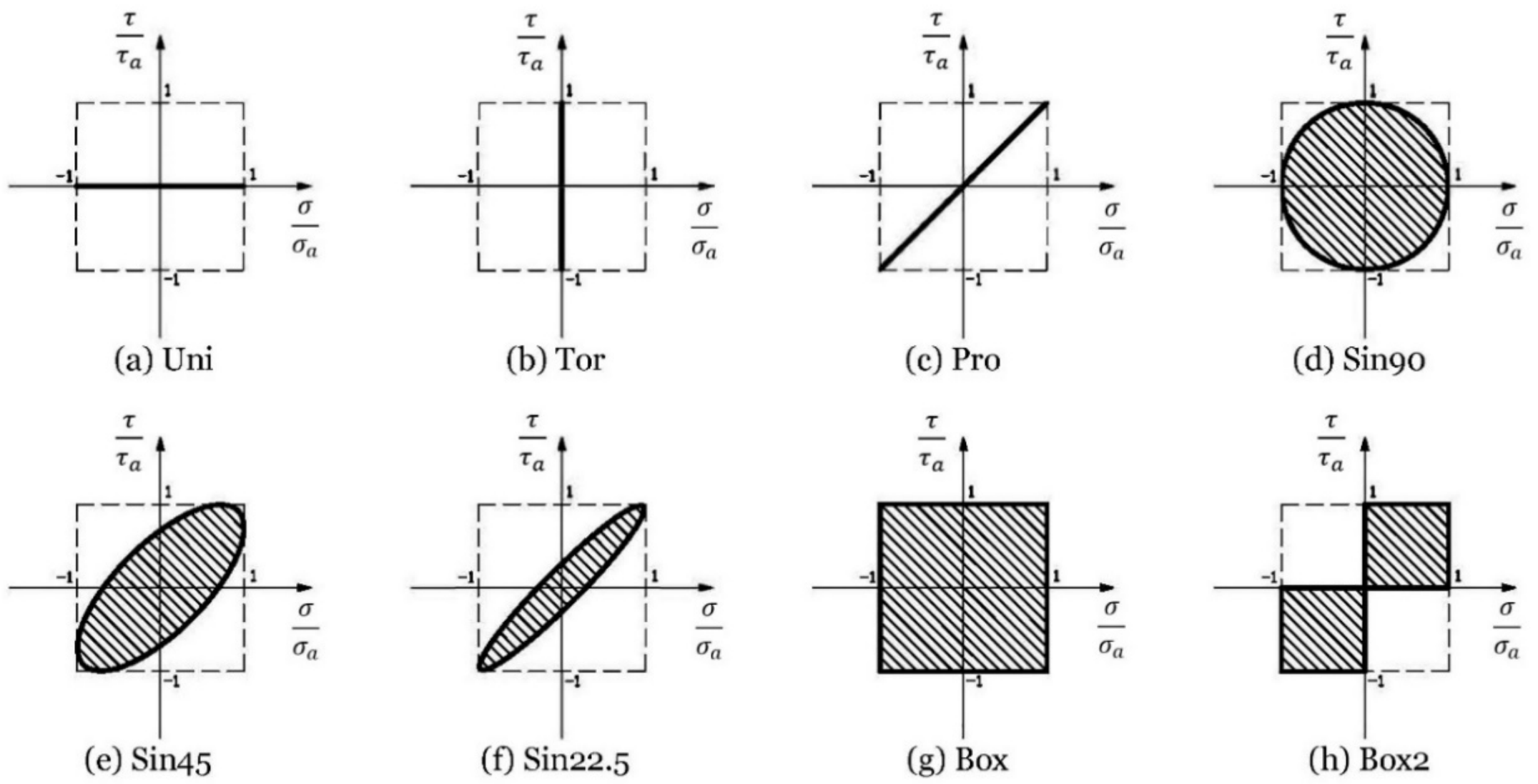

Fig. 5 Load Paths of Collected Data in the Current Study

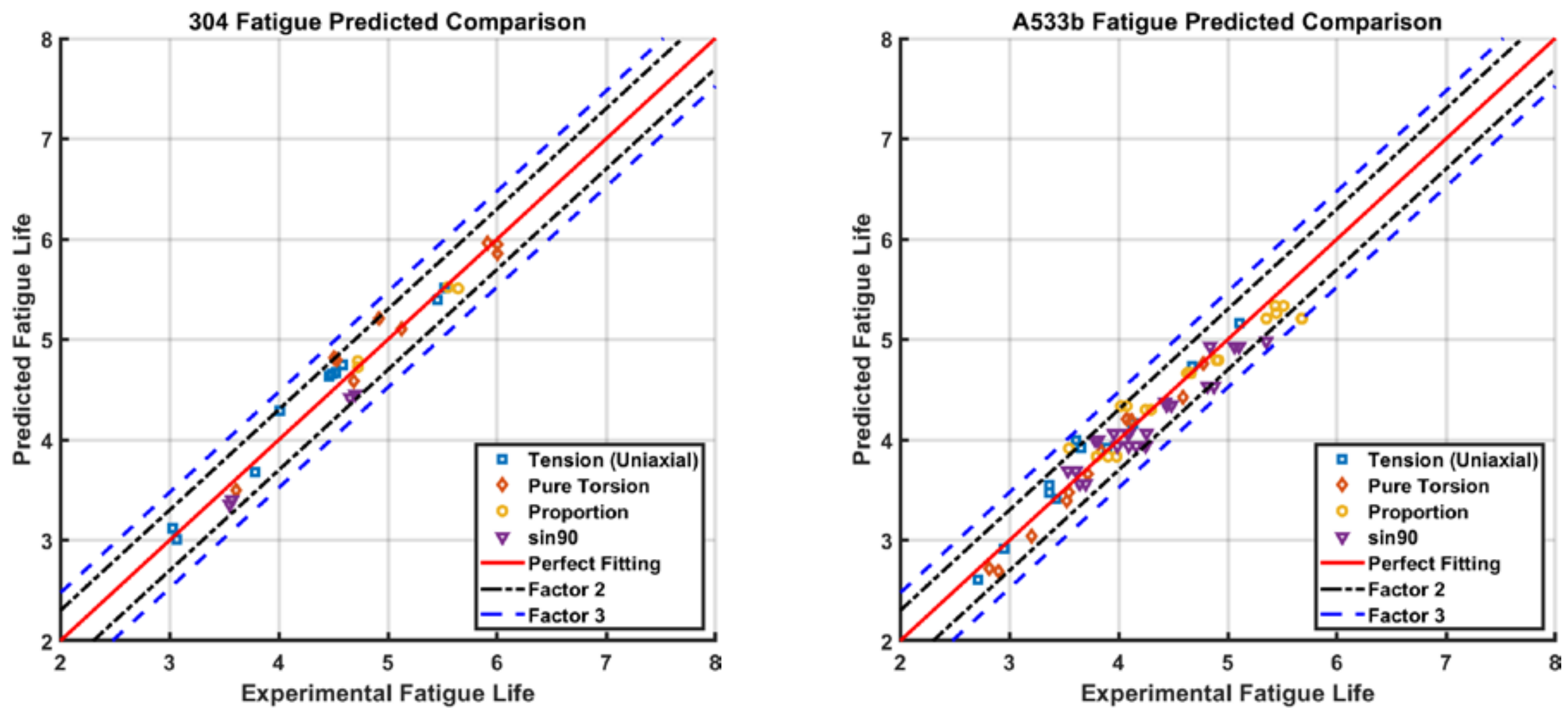

(a) AISI 304 Steel (b) A533B

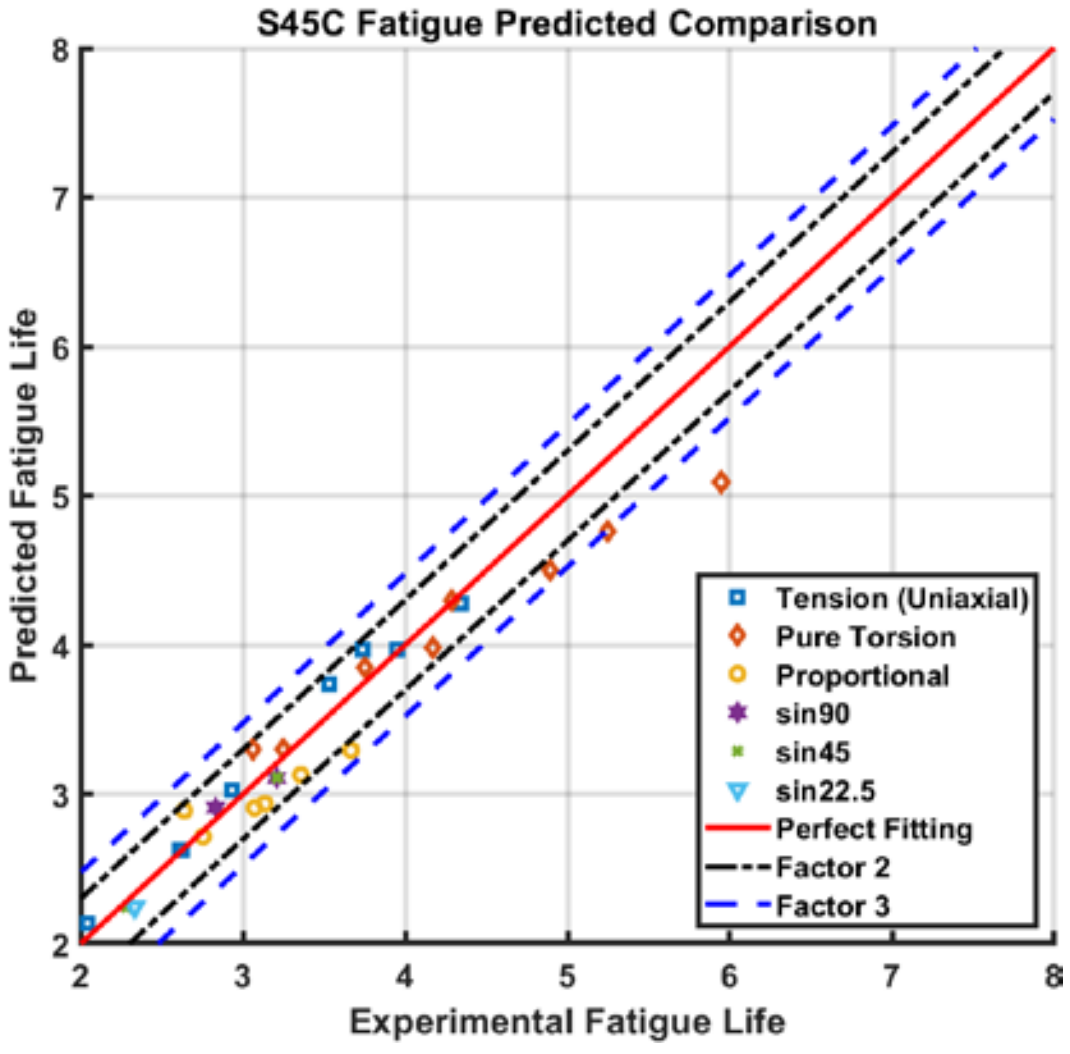

(c) S45C Steel

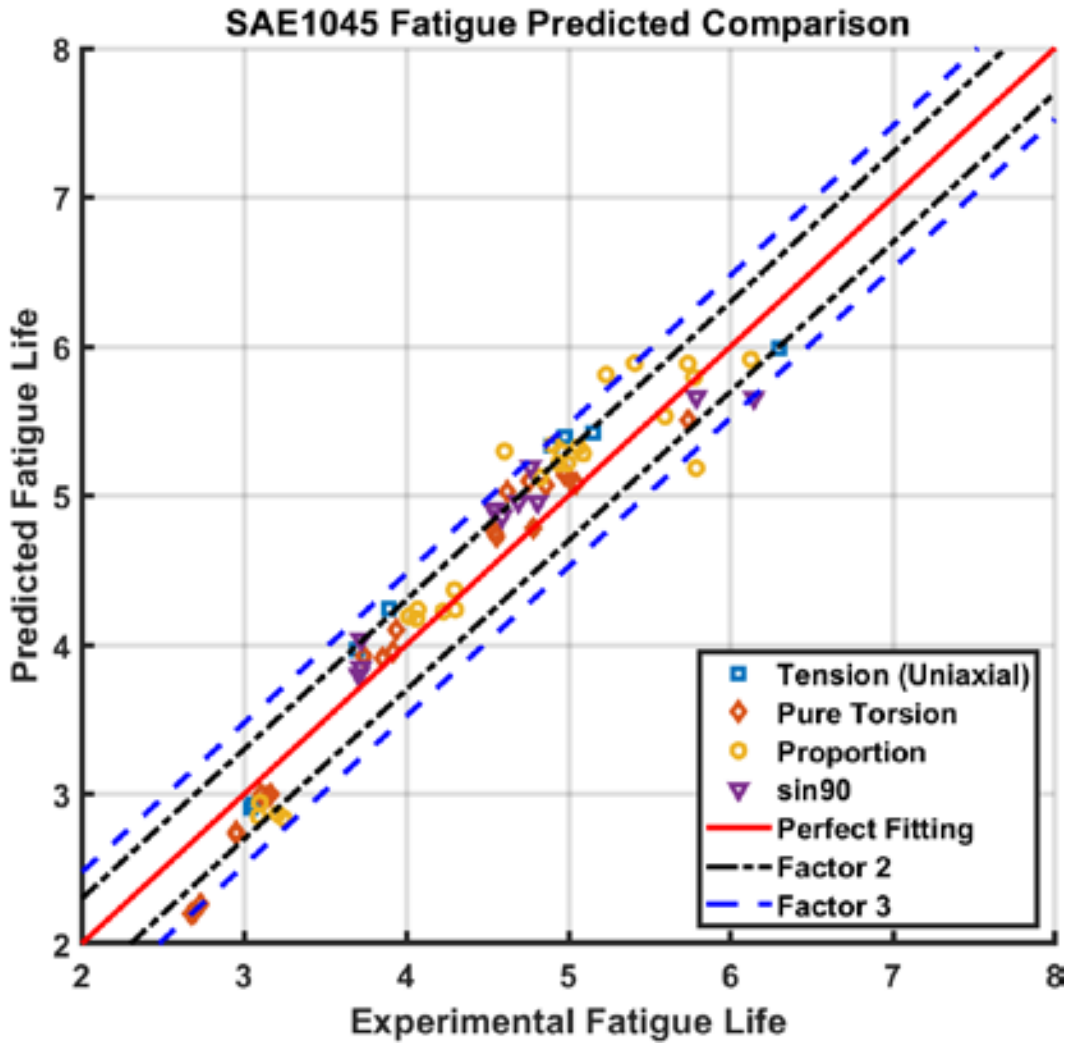

(d) SAE 1045

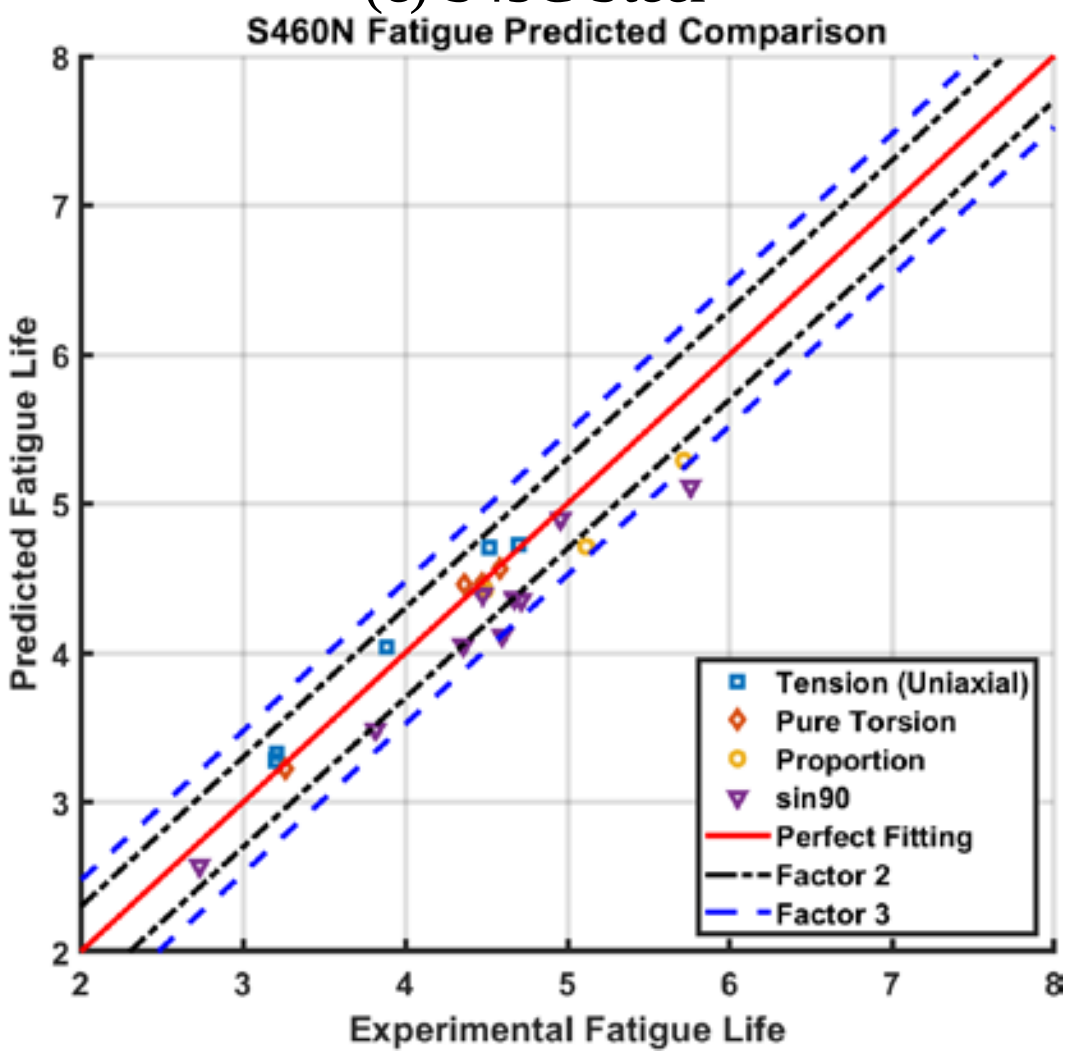

(e) S460N

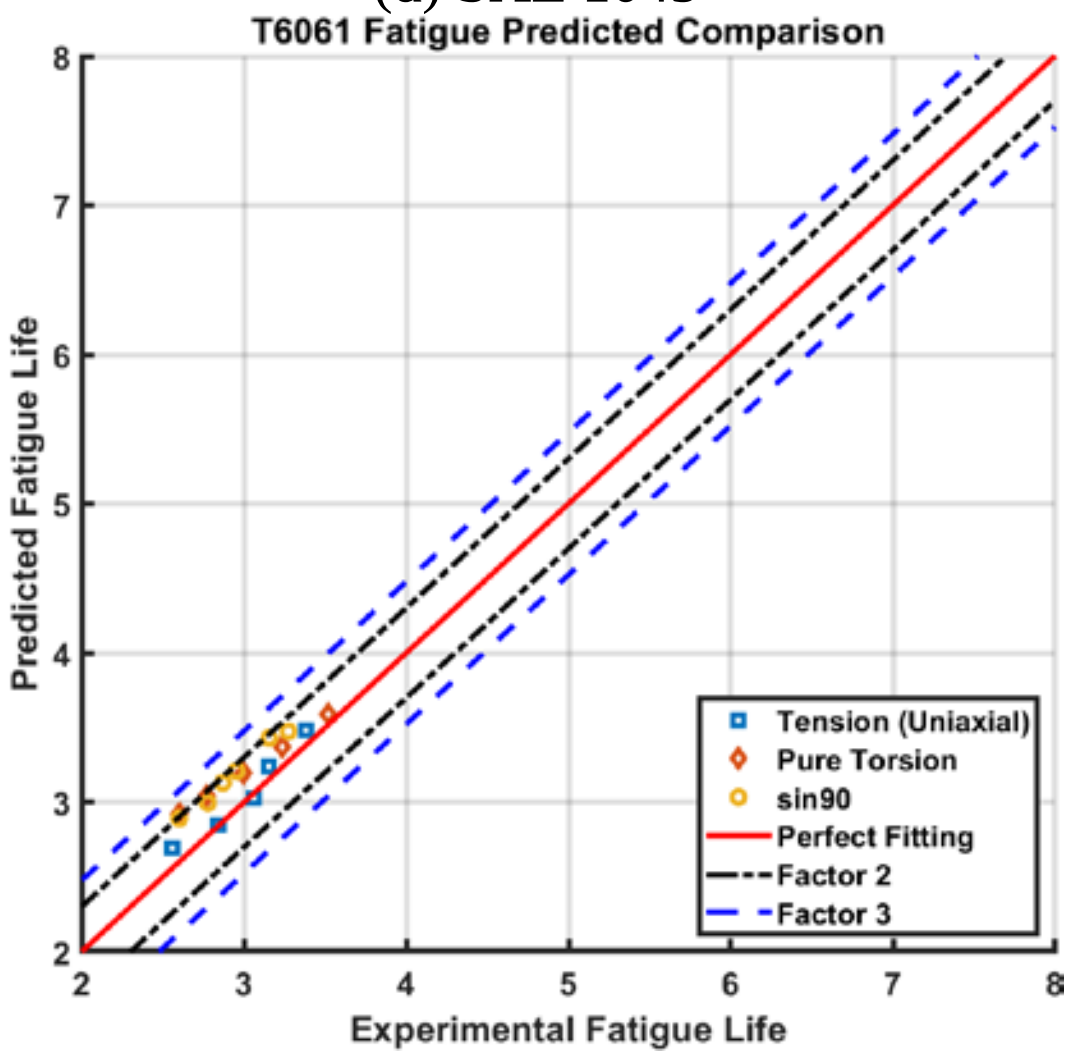

(f) Al T6061

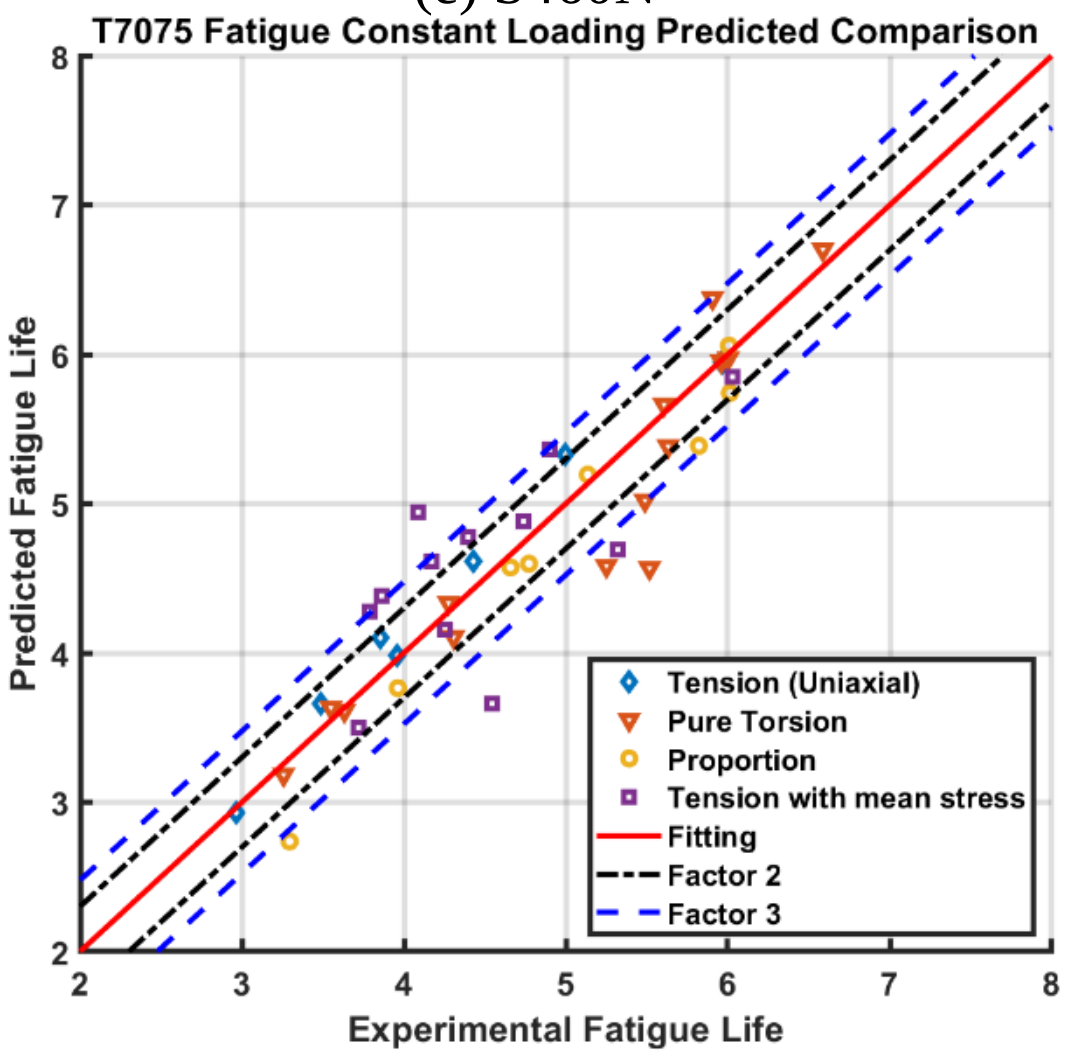

(g) Al T7075

Fig. 6 Predicted Fatigue Life and Experimental Fatigue Life Comparison in Constant Loading Case: (a) AISI 304 Steel (b) SM45C Steel (c) A533B (d) S460N (e) SAE 1045 (f) Al T6061 (g) Al T7075

### 3.2 In-house data generation and model validation

The fatigue specimens are tested using the MTS Landmark servo hydraulic test system. The fatigue test system and experimental setup are shown in Fig. 7. Fatigue loading testing includes both uniaxial and multiaxial spectrum loading conditions.

The 7075-T6 aluminum alloy specimen is selected. The geometries and dimensions of the specimen are shown in Fig. 8. The geometry of the specimen is designed with reference to ASTM E2207-15. The detail of selected specimen and test setup can be found in [2,17].

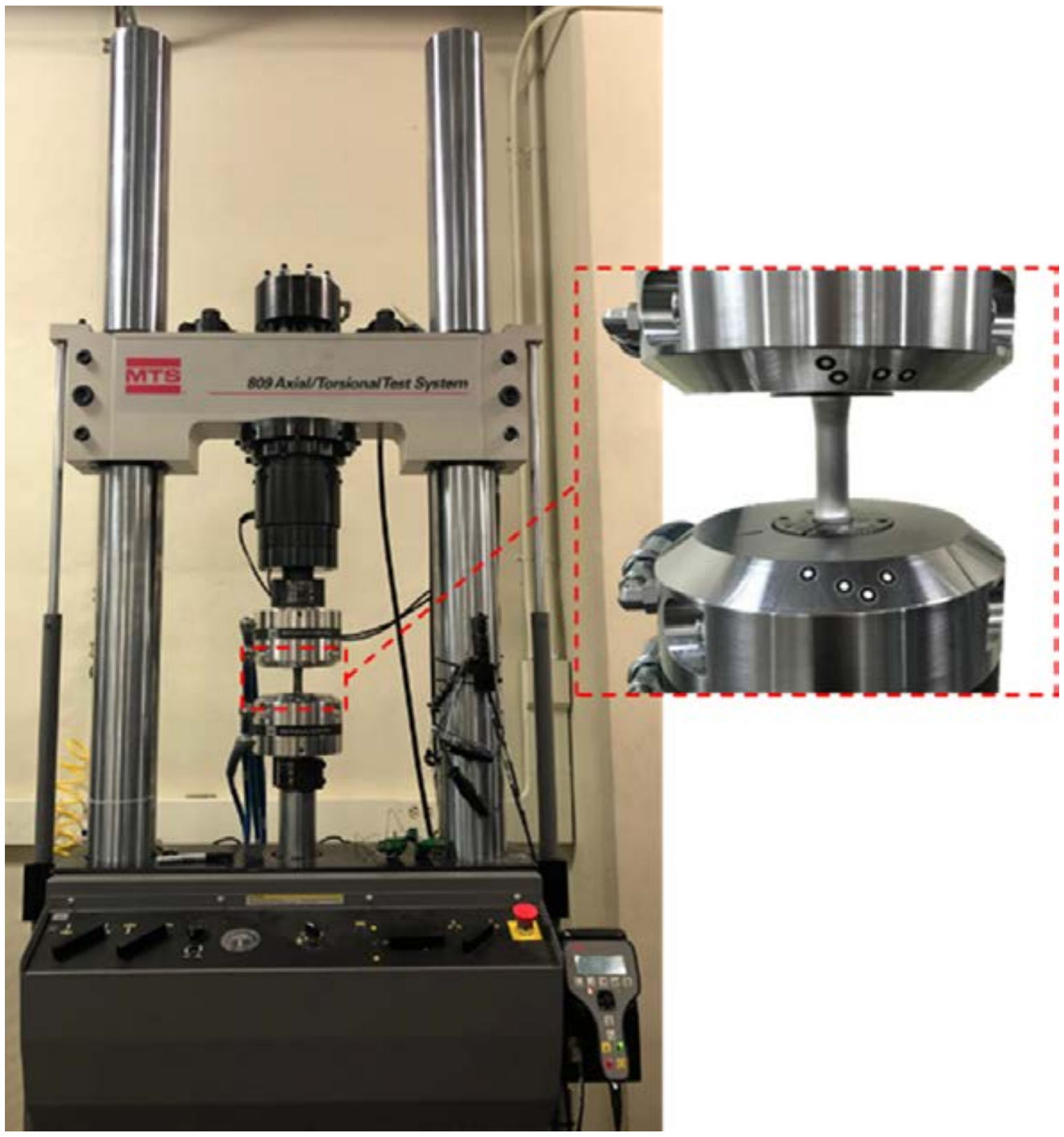

Fig. 7. Fatigue Testing Setup [2]

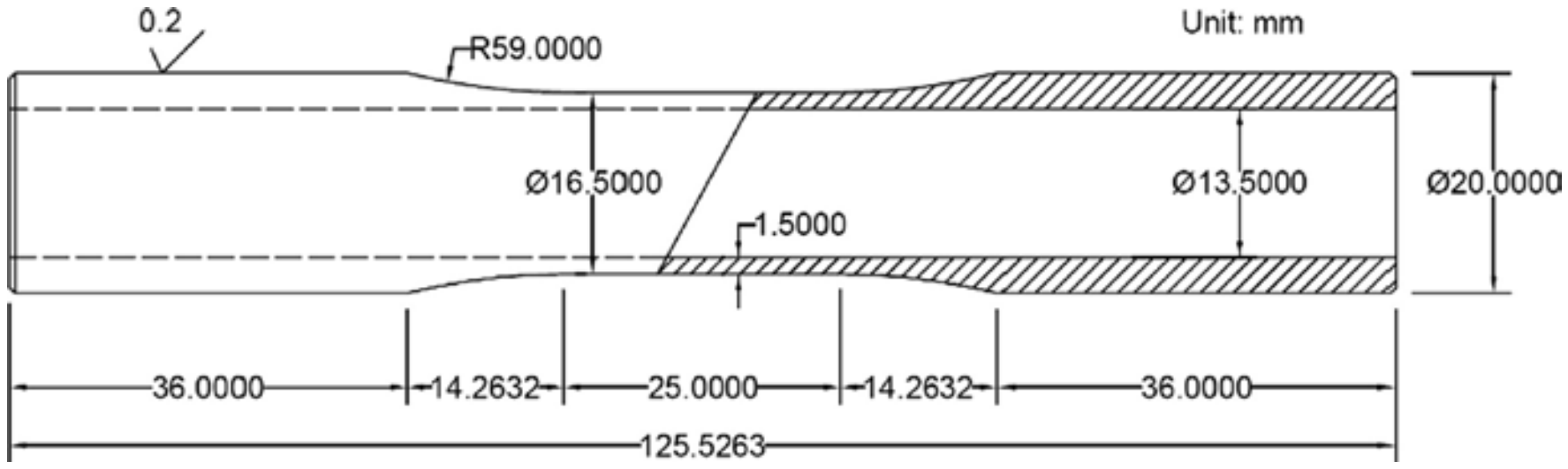

Fig. 8. Specimen Design

Different standard fatigue spectrums were generated. The detail of the equation used for several spectrum generation can be found in [2].

The spectrums for high cycle fatigue (HCF) under uniaxial random loading cases are shown in Fig 9. The spectrums for HCF under multiaxial random loading cases are shown in Fig. 10. The HCF fatigue test results of all specimens are shown in Table 3, which includes the test number, the load spectrum used, and the number of reversals to failure. The spectrums for LCF+HCF under uniaxial and multiaxial random loading cases are shown in Fig. 11. Table 4 summarizes the details of the fatigue testing spectrums under HCF+LCF.

The model predictions are plotted with experimental data in Fig. 12 and excellent prediction results can be observed again for random loadings. This clearly demonstrates the validity of the proposed time-derivative damage model.

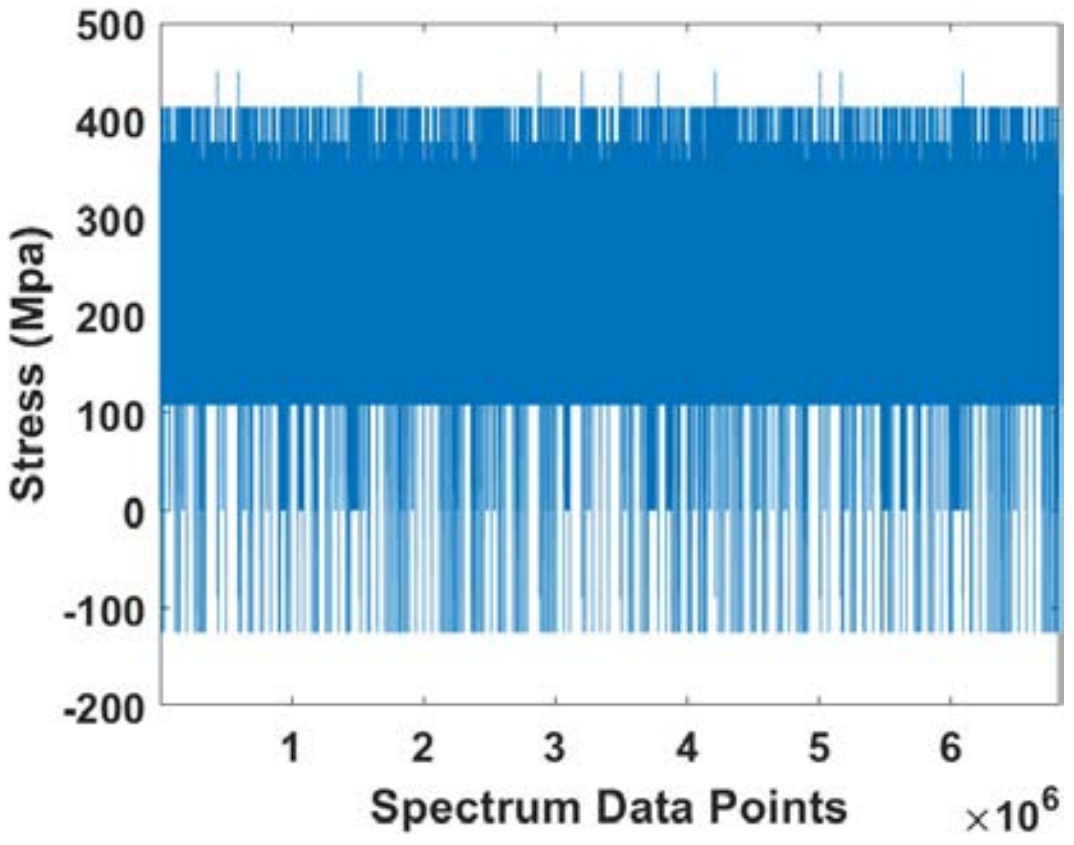

(a) FELIX Spectrum

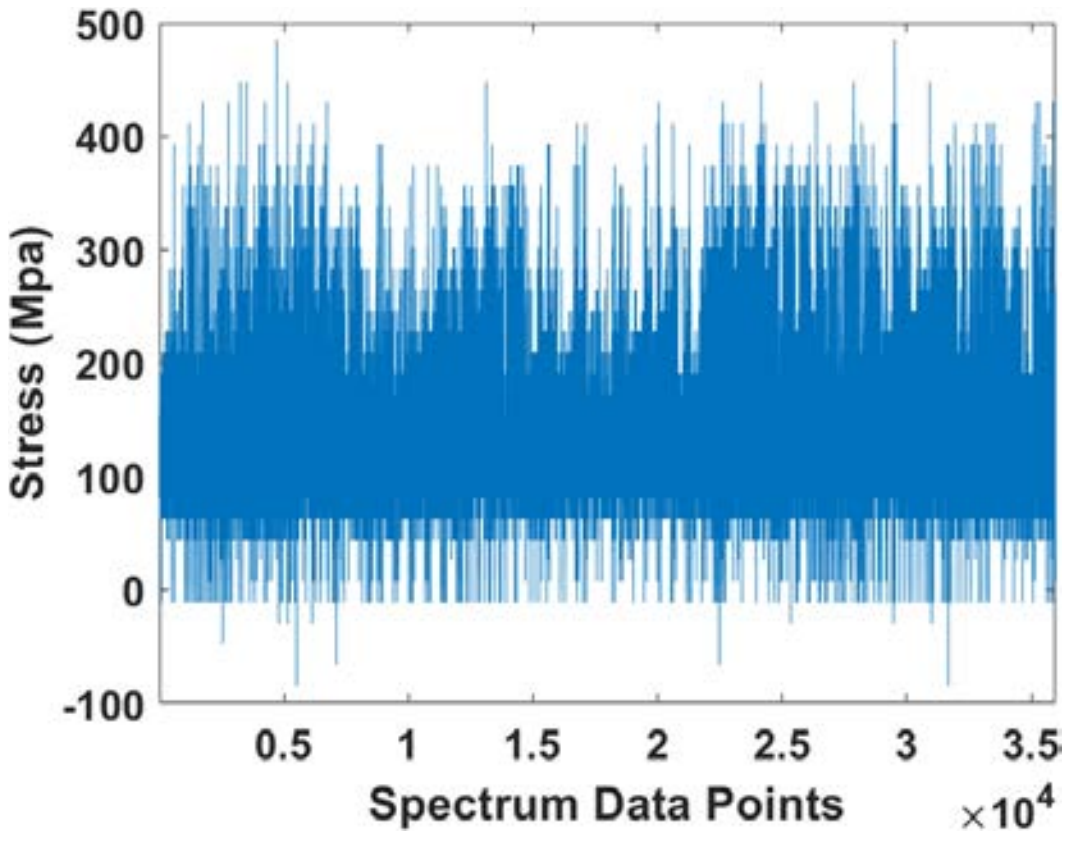

(b) FELIX + 35 Spectrum

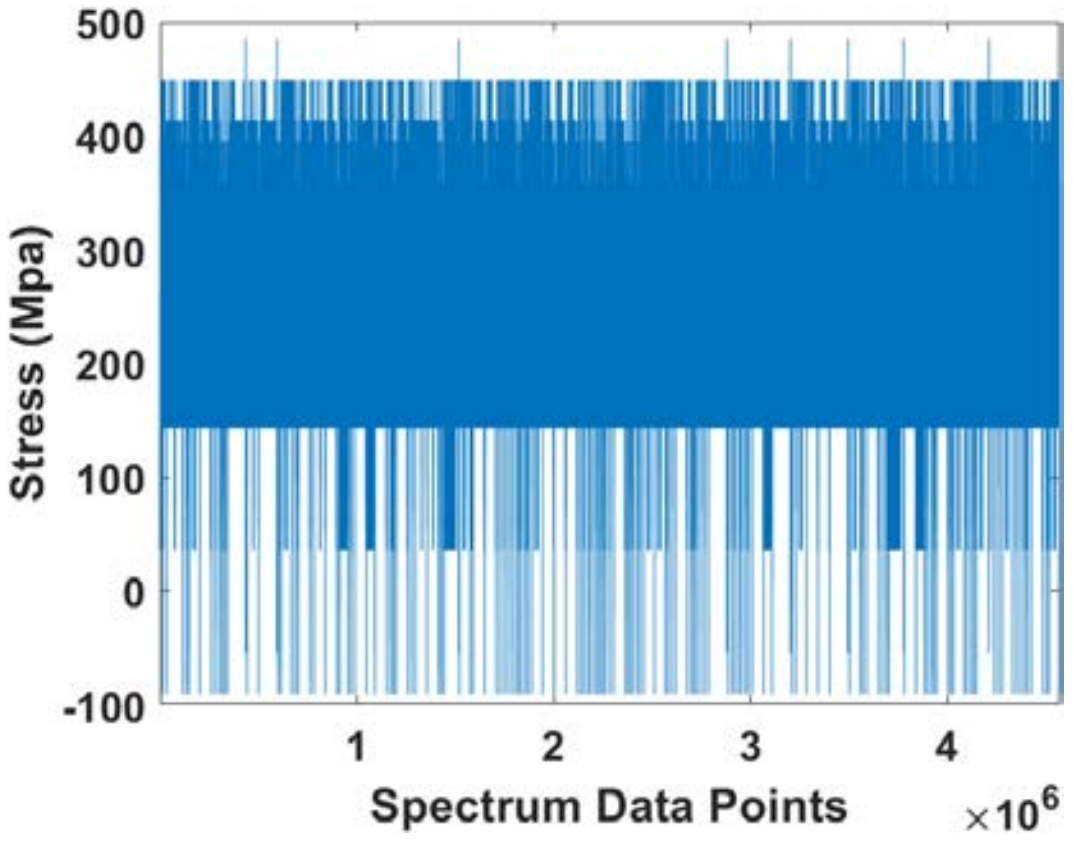

(c) Modified FELIX Spectrum

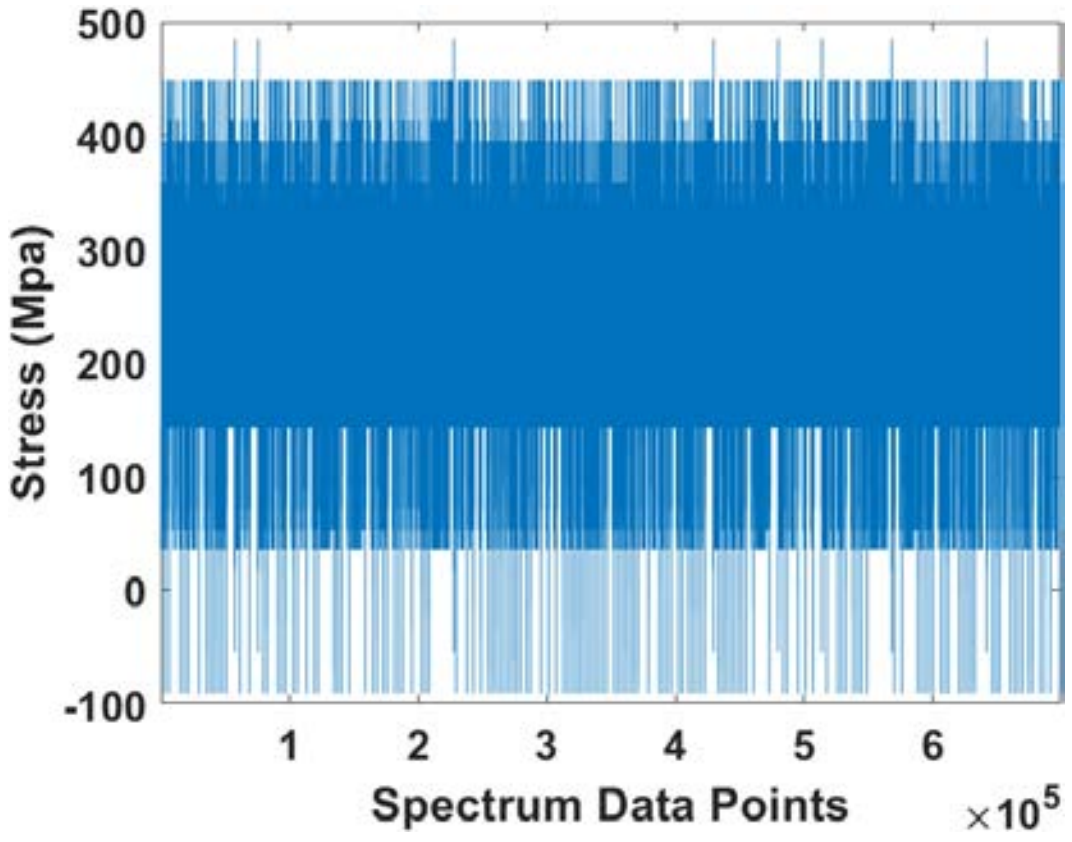

(d) Simplified Max(FELIX) + 35 Spectrum

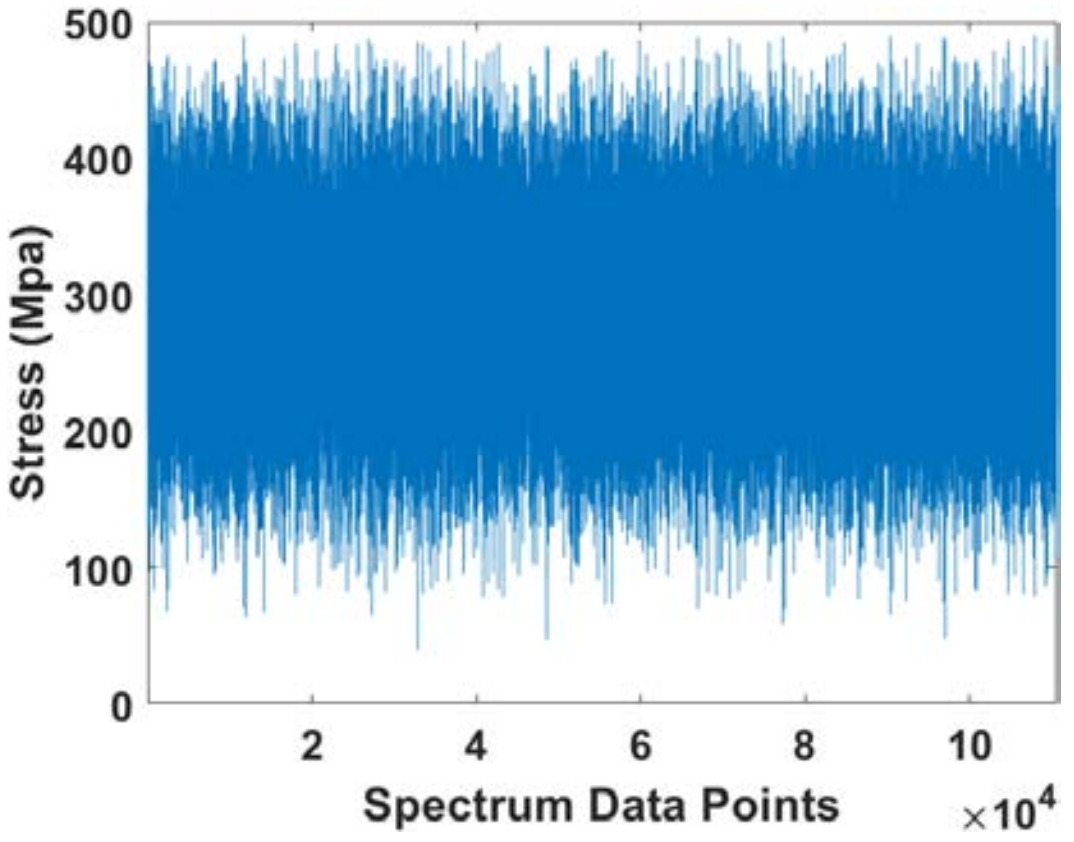

(e) Linear Spectrum

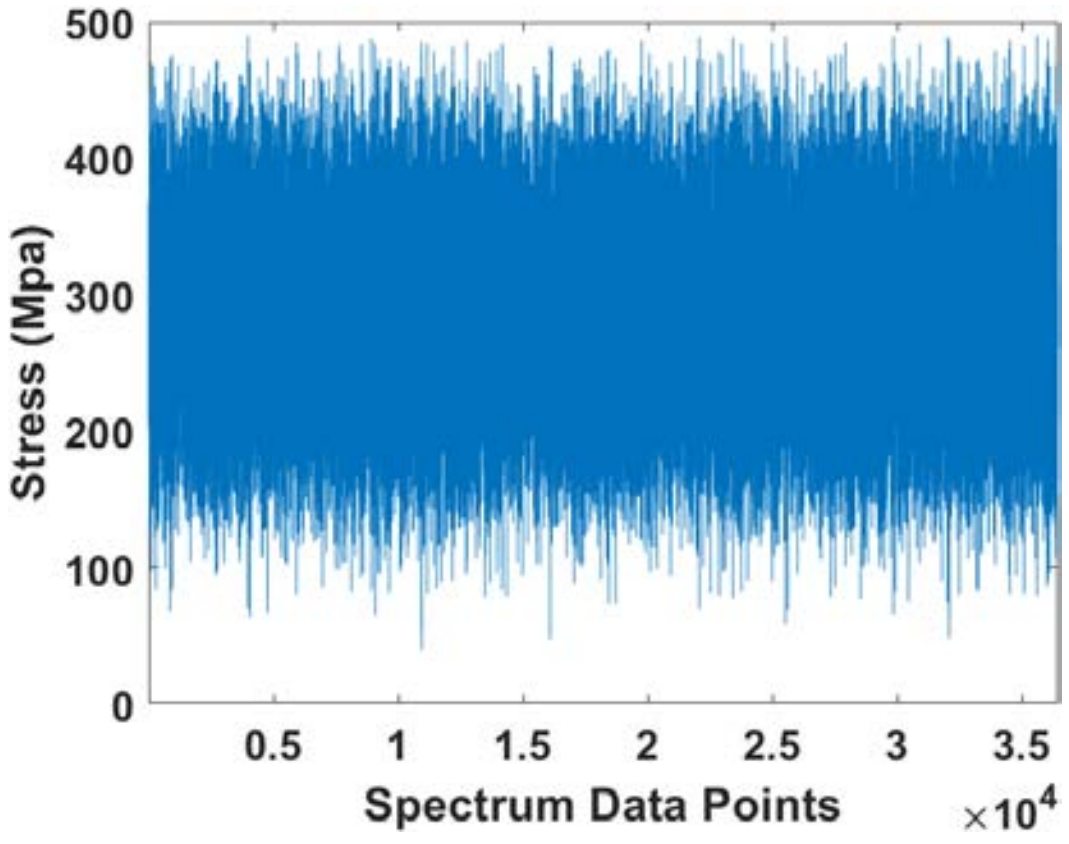

(f) Simplified Linear Spectrum

Fig. 9 Generated Spectrum for Fatigue Testing Under Uniaxial Random Loading

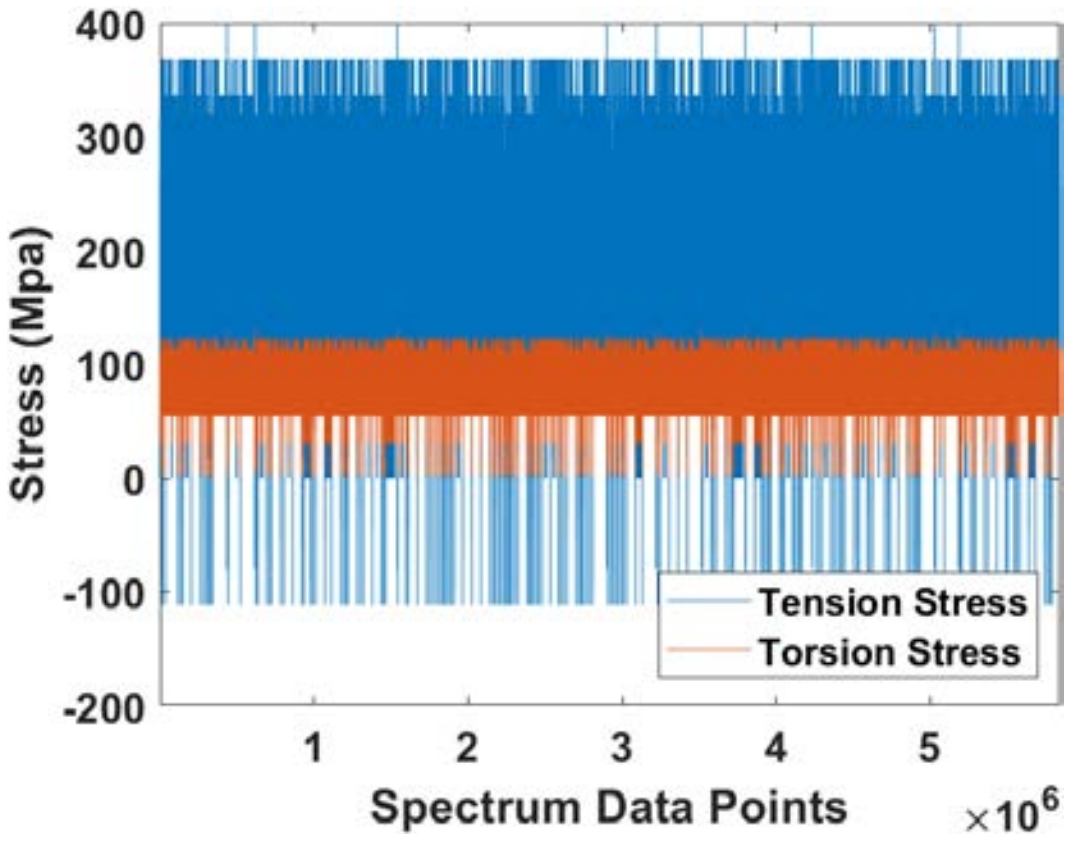

(a) Proportional FELIX Spectrum

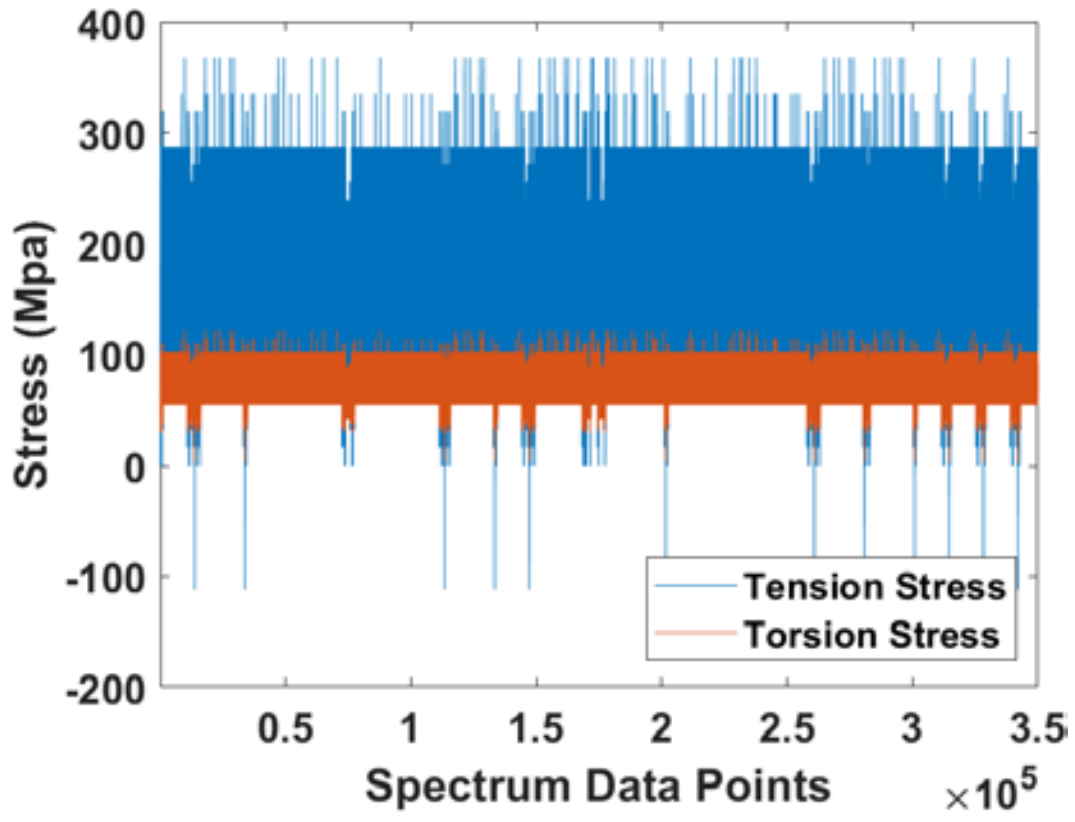

(b) Modified Proportional FELIX Spectrum

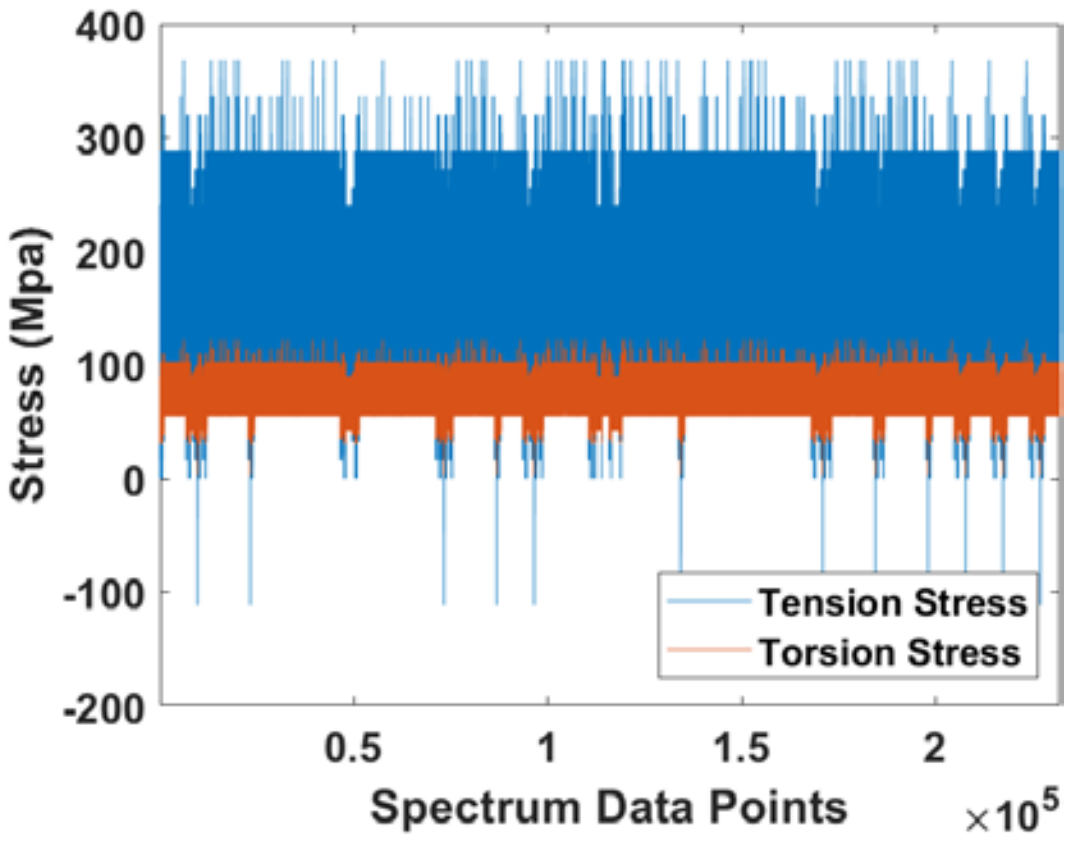

(c) CWT Edited Proportional FELIX Spectrum

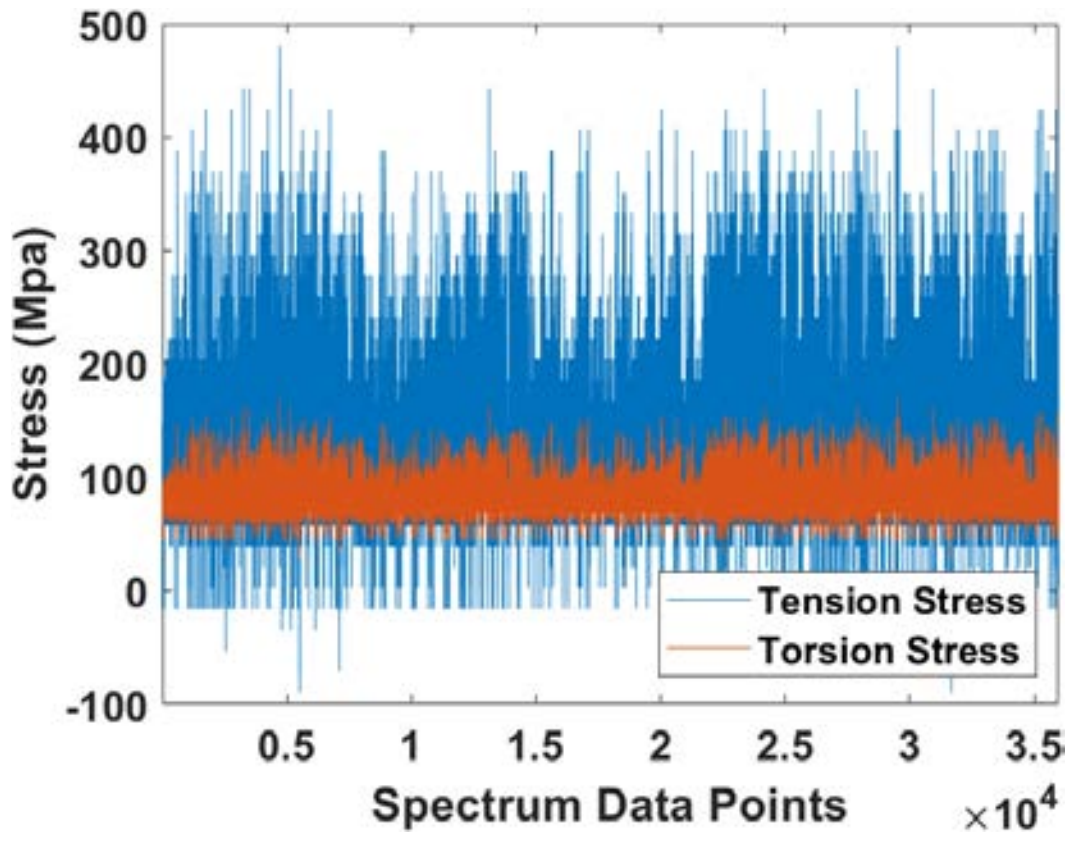

(d) Nonproportional FELIX Spectrum

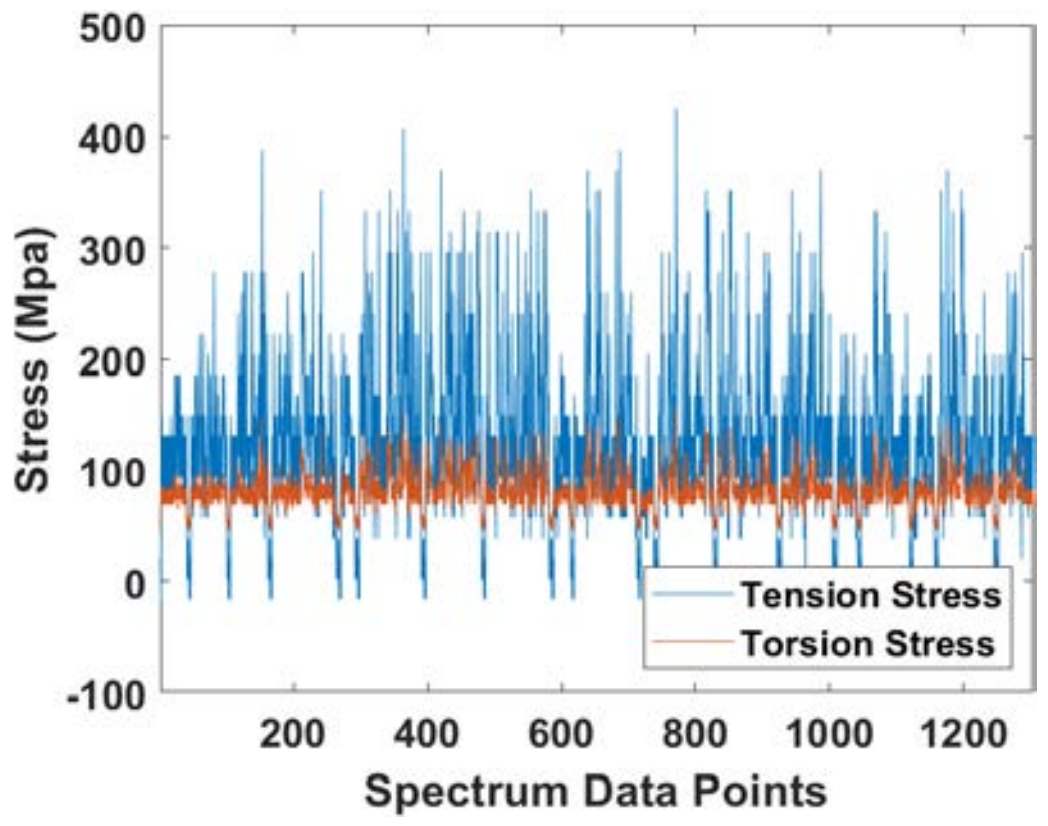

(e) Modified Nonproportional FELIX Spectrum

Fig. 10 Generated Spectrum for Fatigue Testing Under Multiaxial Random Loading

Table 3. Summary of Fatigue Testing Spectrum Under Random Loading Conditions in HCF.

| Spectrum name | Loading path | Fatigue life (number of reversals) |
|---|---|---|
| Linear | Uniaxial | 2,112,921 |
| Simplified linear | Uniaxial | 736,909 |
| FELIX | Uniaxial | 5,956,555 |
| FELIX + 35 | Uniaxial | 1,971920 |
| Modified FELIX | Uniaxial | 2,905,820 |
| Simplified max(FELIX) + 35 | Uniaxial | 1,196,807 |
| Proportional FELIX | Multiaxial | 1,338,786 |
| Modified proportional FELIX | Multiaxial | 1,745,638 |
| CWT edited proportional FELIX | Multiaxial | 5,106,200 |
| Nonproportional FELIX | Multiaxial | 1,191,522 |
| Modified nonproportional FELIX | Multiaxial | 405,053 |

Table 4. Summary of Fatigue Testing Spectrum Under Random Loading Conditions in HCF+LCF.

| Spectrum name | Loading path | Fatigue life (number of reversals) |
|---|---|---|
| Linear_HCF+LCF | Uniaxial | 68,894 |
| Nonlinear_HCF+LCF | Uniaxial | 35,908 |
| Proportional_HCF+LCF | Multiaxial | 11,046 |
| Nonproportional_HCF+LCF | Multiaxial | 17,575 |

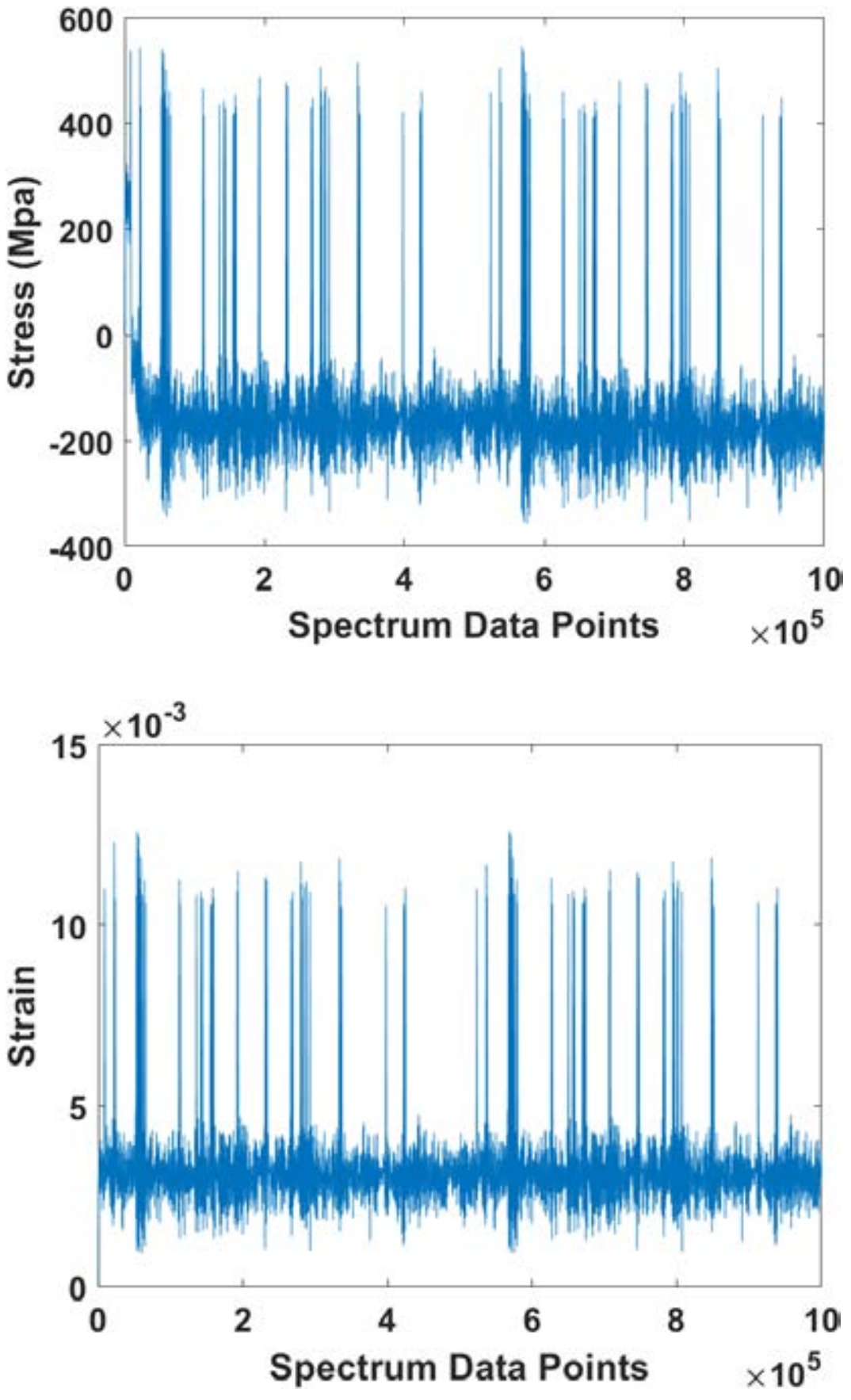

(a) HCF+LCF Linear Spectrum

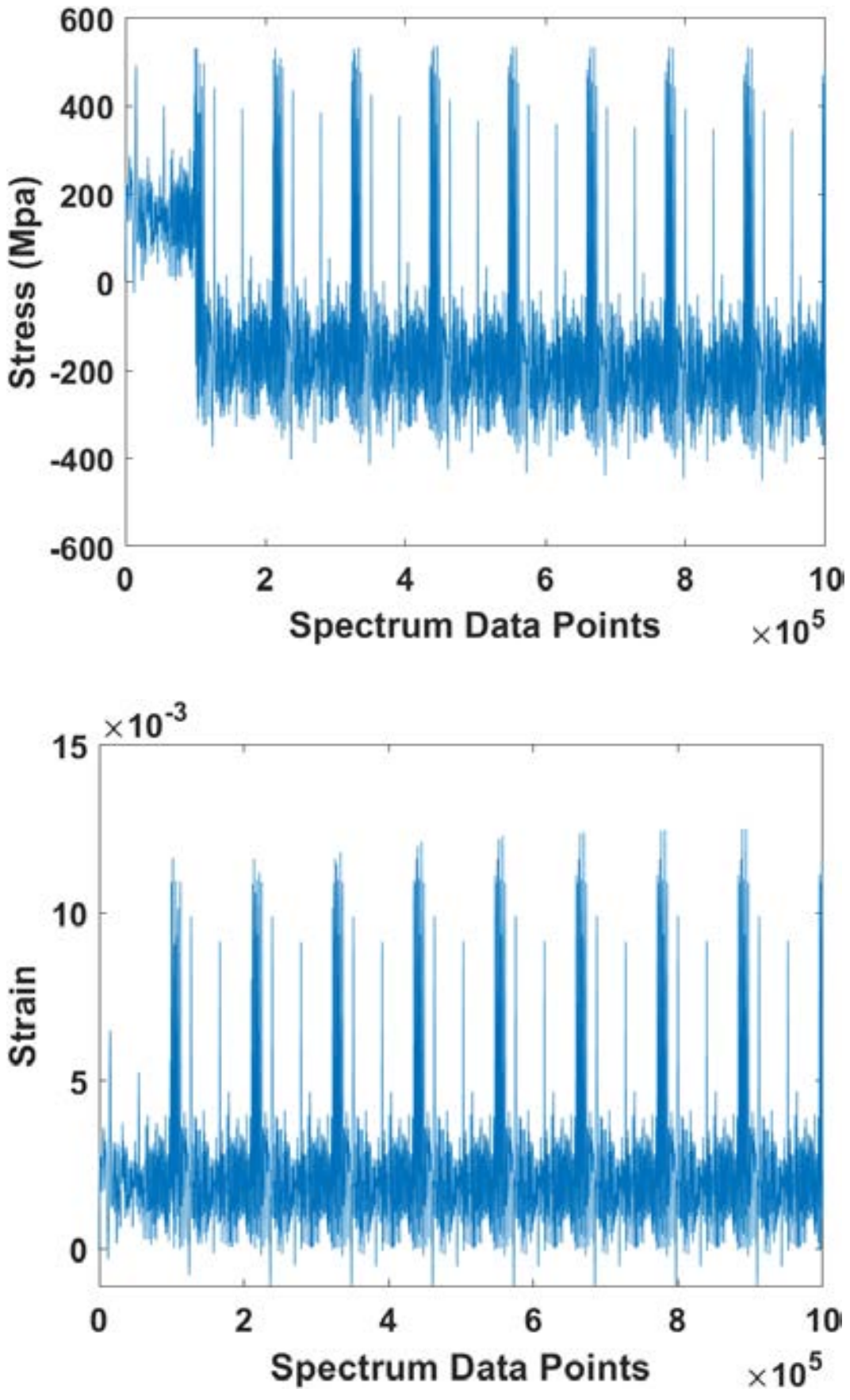

(b) HCF+LCF Nonlinear Spectrum

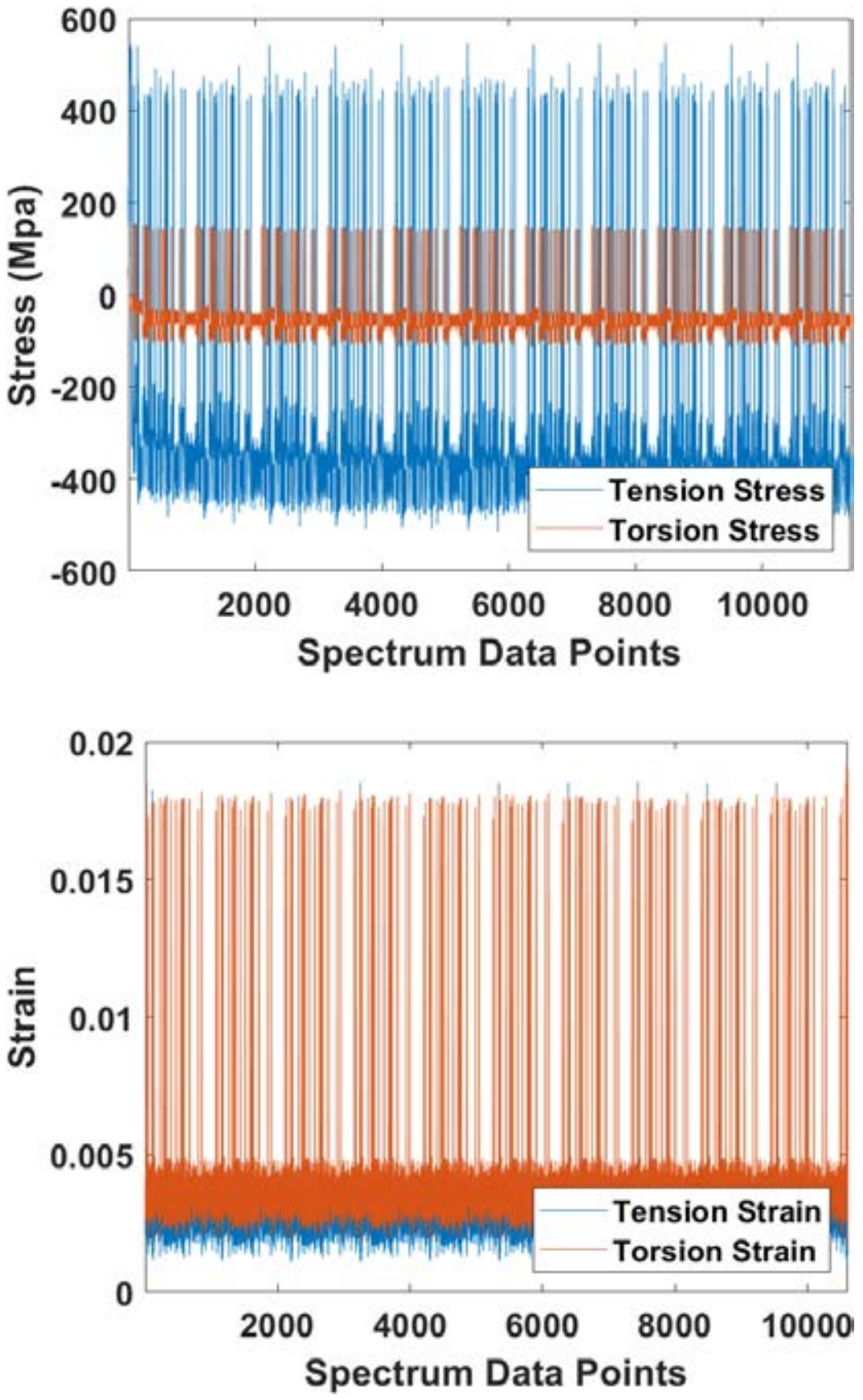

(c) HCF+LCF Proportional Spectrum

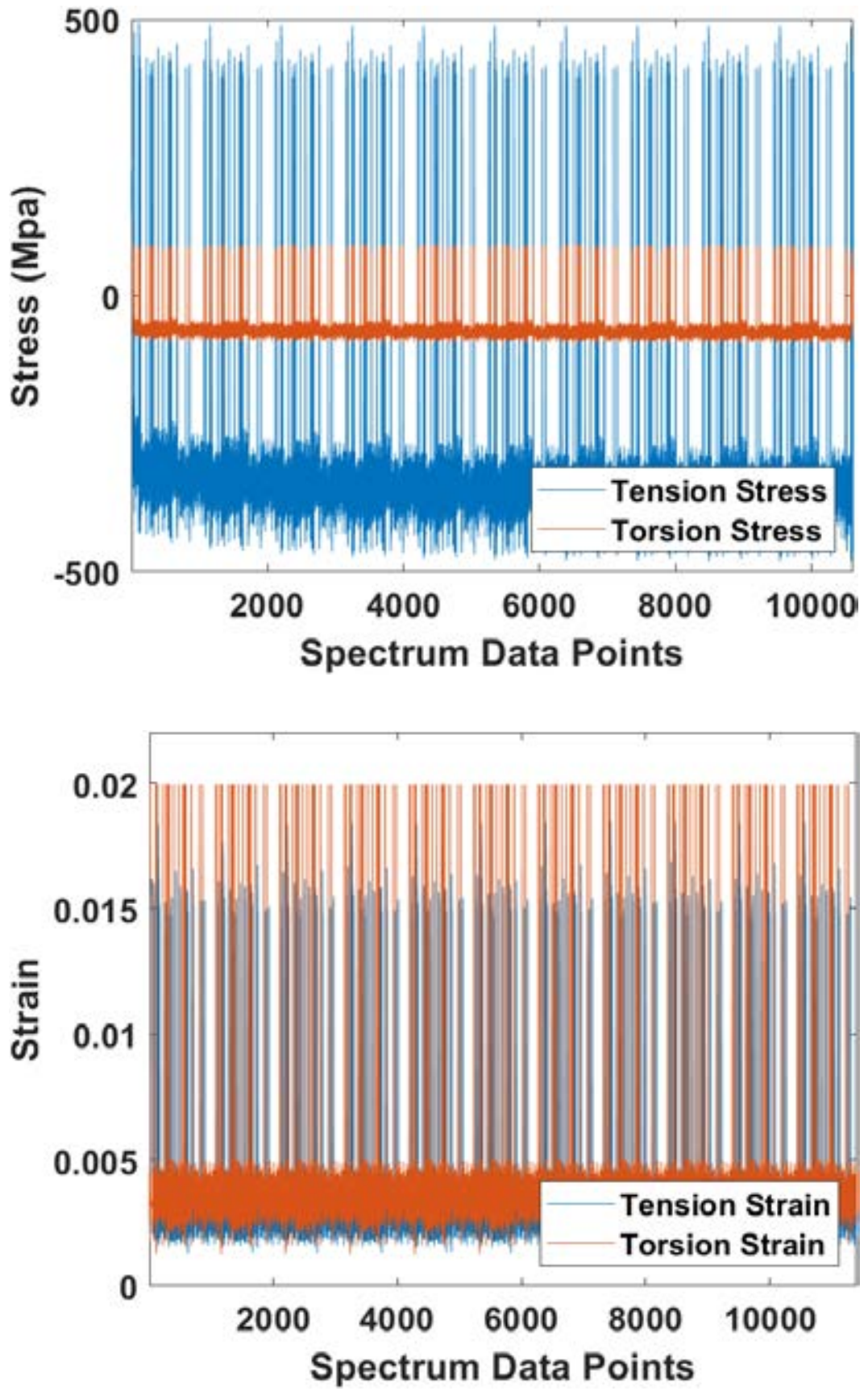

(d) HCF+LCF Proportional Spectrum

Fig. 11 HCF+LCF Spectrum for Fatigue Testing Under Multiaxial Random Loading

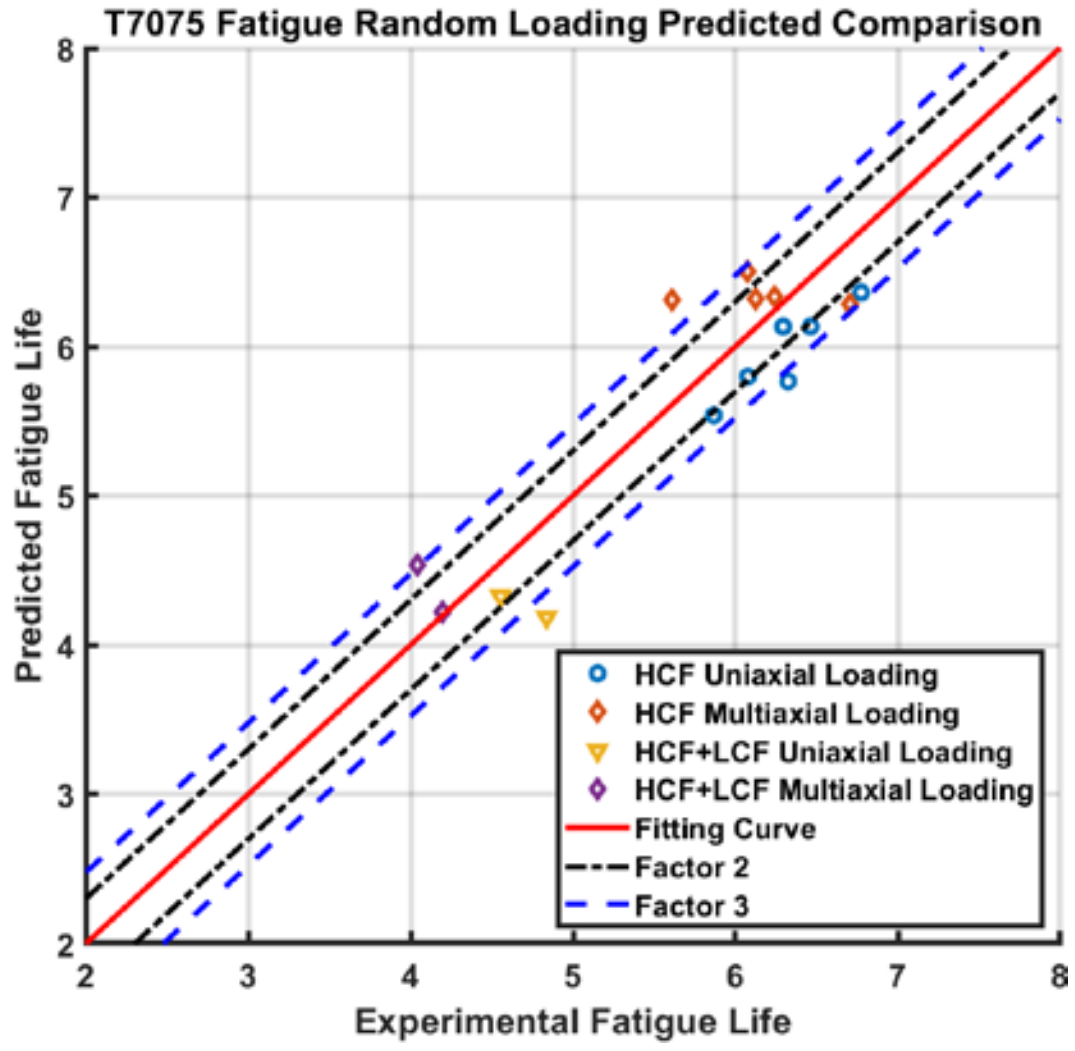

Fig. 12 Predicted Fatigue Life and Experimental Fatigue Life under Random Loading Case for Aluminum T7075-T6

## 4 Conclusion and Future Work

A new energy-based time-derivative damage method is proposed. This model can deal with both constant loading and random loading cases under both HCF and LCF conditions. The model uses the concept based on a new damage accumulation method which is developed in analogy with the fatigue crack growth analysis. The model is validated by extensive experimental data from open literature and in-house testing. The predicted results show excellent agreement a wide range of materials under different loading conditions.

Major conclusions from the proposed study can be summarized below.

1. The theoretical derivation and experimental validation demonstrate the feasibility using a time-derivative formulation for arbitrary loading spectrum life prediction without the need for cycle counting.
2. The time-based model is suitable for dealing both the high cycle fatigue (HCF) and low cycle fatigue (LCF) region since the energy-based method has been used. It appears that the proposed model can automatically account for the plasticity effect by including the stress/strain terms in the energy calculation

3. The proposed damage definition is not using the cycle/cycle fraction and is directly linked with the energy and is analogy to crack. It appears a power law rate function similar the classical crack growth function is applicable to a wide range of metallic materials.
4. There is no linear or nonlinear damage accumulation separation in the proposed method, unlike classical fatigue damage accumulation models. Direct time-domain integration or load-history integration is used. This methodology offers a new, systematic, and alternative way for fatigue life prediction with no restriction for the type and form of external loading.

Several future research directions and limitations of the proposed study are listed below. First, the proposed study makes the hypothesis for the damage definition and rate function format. It is interesting to see the mechanism explanation and support for this hypothesis using advanced experimental testing and imaging analysis. For example, in situ SEM/TEM testing may be able to reveal the micro-voids and dislocation evaluation which may show correlation with the proposed damage variables. This will need significant future theoretical and experimental studies. Next, the proposed study aims for fatigue analysis and replaces the time domain integration with load-path integration and ignores the rate-dependent behavior, such as creep, corrosion, and oxidization. Extension to creep-fatigue, corrosion-fatigue, and other rate-dependent fatigue behavior will need further study. It is expected that the time-dependent integration will be required for the frequency effect and dwell-time effect typically seen in corrosion-fatigue and creep-fatigue problems. Last but not least, the uncertainties associated with the fatigue crack initiation is not addressed here and only the mean behavior is considered. Probabilistic modeling integrating with the proposed methodology will need further investigation, especially when both material and loading uncertainties are included. The proposed methodology offers a unique capability for integration with direct time-domain random process theories, such as principal component analysis or Kahunen-Leove expansion techniques.

**Acknowledgements**

The research is partially supported by fund from NAVAIR through subcontract from Technical Data Analysis, Inc (TDA) (contract No. N68-335-18-C-0748, program manager: Krishan Goel). The support is greatly appreciated.